\documentclass[12pt]{iopart}
\usepackage[utf8]{inputenc}  
\usepackage[T1]{fontenc}  
\usepackage[english]{babel} % Should enable hyphenation
\usepackage{epsfig} % to include .eps figures
\usepackage{parskip}
\usepackage{setspace}
\usepackage{subcaption}
\usepackage{ifpdf}
\usepackage{nomencl}
\DeclareMathSizes{10}{10}{10}{10}
\begin{document}
%\ifpdf
%    \graphicspath{{Figs/PNG/}{Figs/PDF/}{Figs/}}
%\else
%    \graphicspath{{Figs/EPS/}{Figs/}}
%\fi
%\ \\

\title[Maximal near-field radiative heat transfer between two plates]{Maximal near-field radiative heat transfer between two plates}

\author{Elyes NEFZAOUI, Younès EZZAHRI, Jérémie DREVILLON and Karl JOULAIN}

\address{Institut Pprime, CNRS-Universit\'e de Poitiers-ENSMA, D\'epartement Fluides, Thermique, Combustion, ENSIP-B\^atiment de m\'ecanique, 2, Rue Pierre Brousse, F 86022 Poitiers, Cedex, France}
\ead{elyes.nefzaoui@univ-poitiers.fr}
\begin{abstract}
A parametric study of Drude and Lorentz models performances in maximizing near-field radiative heat transfer between two semi-infinite planes separated by nanometric distances at room temperature is presented in this paper. Optimal parameters of these models that provide optical properties maximizing the radiative heat flux are reported and compared to real materials usually considered in similar studies, silicon carbide and heavily doped silicon in this case. Results are obtained by exact and approximate (in the extreme near-field regime and the electrostatic limit hypothesis) calculations. The two methods are compared in terms of accuracy and CPU resources consumption. Their differences are explained according to a mesoscopic description of near-field radiative heat transfer. Finally, the frequently assumed hypothesis which states a maximal radiative heat transfer when the two semi-infinite planes are of identical materials is numerically confirmed. Its subsequent practical constraints are then discussed.
\end{abstract}
\maketitle
%\begin{multicols}{2}
\section{Introduction}
It has been shown in the late 1960 \cite{Polder1971,Cravalho1969} that the radiative heat flux (RHF) exchanged by two media in the near-field (NF), i.e. when these media are separated by very small distances (smaller than the thermal radiation characteristic wavelength $\lambda_T = \frac{h c}{k_b T}$) could exceed by several orders of magnitude the black body limit. This topic has then received an increasing attention until its recent experimental verifications \cite{Rousseau2009b,Kittel2005,Narayanaswamy2008}. Experiments exclusively focused on asymmetric configurations such as plane-tip or plane-sphere configurations.   
On the other hand, the symmetric plane-plane configuration, potentially useful for various applications such as the cooling of high flux density electronic devices \cite{Guha2012} or thermo-photovoltaic (TPV) conversion of radiative energy \cite{Basu2009bRev}, has been thoroughly investigated from a theoretical point of view by several groups \cite{Francoeur2011,Rousseau2012,Basu2010a}. These theoretical works mainly addressed  dielectrics, usually silicon carbide (SiC) \cite{Francoeur2010,Rousseau2012} which surface phonon-polaritons highly contribute to the NF RHF increase. They also considered materials which support plasmon-polaritons in the wavelength range of thermal radiation at room temperature such as tungsten \cite{Laroche2006,Francoeur2011} or heavily doped silicon (HD-Si) \cite{Basu2010a,Rousseau2009}.\\
In the present numerical work, hypothetical materials modeled by local Drude and Lorentz models are considered.
The aim of this work is to find the sets of parameters of these models that possibly maximize the RHF between two semi-infinite planes of identical materials separated by a nonometric gap at room temperature.
For this purpose, we calculate the exchanged RHF between the two media while varying the different parameters in a wide range. Exact and approximate calculations are performed and their accuracy/resources consumption ratio compared.
Then, the optimal hypothetical material performances are compared to those of usually considered materials, SiC and HD-Si for instance.
Finally, the influence of small discrepancies between the optical properties of the two planes on the exchanged RHF is discussed.
% ------------------------------------------------------------------------
% Formalism
% ------------------------------------------------------------------------
\section{Formalism}
\label{sec:Chap5-formalisms}
The two methods used to calculate NF RHF between two semi-infinite planes and to obtain results presented later in this paper are briefly reminded in this section.
% ------------------------------------------------------------------------
% Calcul Direct : PvH
% ------------------------------------------------------------------------
\subsection{Exact calculation}
\label{sec:Chap5-FullFlux}
~~\\
\begin{figure}[ht]
\begin{center}
\includegraphics[width=0.4\textwidth]{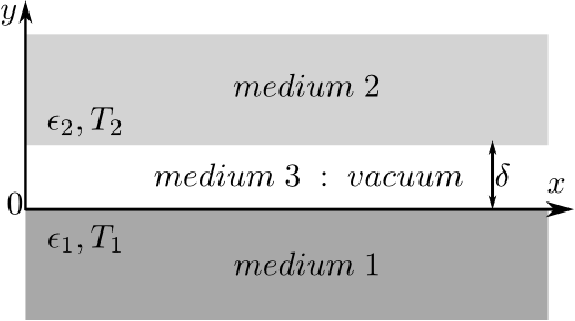}
\caption[]{Two semi-infinite planes separated by a distance $\delta$.}
\label{fig:Chap5-NF-PlanePlane}
\end{center}
\end{figure}
Consider two semi-infinite planes $1$ and $2$ separated by a gap of thickness $\delta$ (Figure \ref{fig:Chap5-NF-PlanePlane}) and characterized by their dielectric functions and temperatures ($\epsilon_1,T_1$) and ($\epsilon_2,T_2$) respectively. The total RHF density exchanged by the two media is given by \cite{Volokitin2007}:
\nomenclature[G]{$\epsilon$}{Dielectric permittivity}%
\nomenclature[S]{$j$}{Medium number, $j \in \{1,2\}$}%
\begin{eqnarray}
\dot{q} & = & \dot{q}_{prop} + \dot{q}_{evan}
\label{eq:Chap5-QPropEvan}
\end{eqnarray}
where
\begin{eqnarray}
\fl \dot{q}_{prop} & = & \sum_{i=s,p} \int_0^{\infty} \frac{d \omega}{2 \pi} 
[ \Theta(\omega,T_1)-\Theta(\omega,T_2)] 
\int_0^{\frac{\omega}{c}} \frac{d^2 q}{(2 \pi)^2} 
\frac{(1-|r^i_{31}|^2)(1-|r^i_{32}|^2)}{|1-r_{31}^i r_{32}^i e^{2 \imath \gamma_3 \delta} |^2}
\label{eq:Chap5-QpropVolokitin}
\end{eqnarray}
\nomenclature[S]{$prop$}{Propagative waves contribution}%
and
\begin{eqnarray}
\fl \dot{q}_{evan} & = & \sum_{i=s,p} 4 \int_0^{\infty} \frac{d \omega}{2 \pi} 
[\Theta(\omega,T_1)-\Theta(\omega,T_2)] 
\int_{\frac{\omega}{c}}^{\infty} \frac{d^2 q}{(2 \pi)^2} 
e^{2  \imath \gamma_3 \delta} 
\frac{{Im}(r_{31}^i) {Im}(r_{32}^i)}{|1-r_{31}^i r_{32}^i e^{2 \imath \gamma_3 \delta} |^2}
\label{eq:Chap5-QevanVolokitin}
\end{eqnarray}
\nomenclature[S]{$evan$}{Evanescent waves contribution}%
are the contributions of propagative and evanescent waves respectively.
\begin{eqnarray}
\Theta(\omega,T) & = & \frac{\hbar \omega}{e^{\frac{\hbar \omega}{k T}}-1}
\label{eq:Chap5-PlanckOscillatorEnergy}
\end{eqnarray}
\nomenclature[G]{$\Theta$}{Planck oscillator mean energy}%
is the mean energy of a Planck oscillator at a temperature $T$.
$r_{3j}^i$ are Fresnel reflection coefficients for an $i$-polarized wave ($i \in \{s,p\}$) propagating from medium $3$ to medium $j$.
\nomenclature[L]{$r_{3j}$}{Fresnel reflexion coefficient between media $3$ and $j$}%
\nomenclature[U]{$i$}{Wave polarization, $i \in \{s,p\}$}%
$\gamma_3$ is the wave vector normal component in medium $3$ and is given by:
\begin{eqnarray}
\gamma_3 & = & \sqrt{\left( \frac{\omega}{c} \right)^2 - q^2}
\label{eq:Chap5-PlanckOscillatorEnergy}
\end{eqnarray}
\nomenclature[G]{$\gamma$}{Normal wave vector component}%
where $\mathbf{q}$ is the component of the wave vector parallel to the interfaces.
\nomenclature[L]{$\mathbf{q}$}{Transverse wave vector component}%
Expressions \ref{eq:Chap5-QpropVolokitin} and \ref{eq:Chap5-QevanVolokitin} express the total heat flux as the sum of the energy of different existing oscillators at a temperature $T$, transported by different modes ($\omega,\mathbf{q}$).
It is worth noting that for $q > \frac{\omega}{c}$, $\gamma_3$ is imaginary, the corresponding waves are evanescent and their magnitude decreases when going away from the surface. Corresponding modes are surface waves modes.\\
To obtain the total heat flux, a double integration over all modes $(\omega,\mathbf{q})$ is to be made. Its calculation may prove to be very resource-consuming since the cutoff wave vector $q_c$ for the integral over $q$ is not known \textit{a priori}. 
Different authors have proposed different approximations for the cutoff wave vector : $q_c = 1/\delta$ \cite{Rousseau2009,Biehs2010}, $q_c = \sqrt{4/\delta^2 + (\omega/c)^2}$ \cite{benabdallah2010} and $q_c = 1/a$ \cite{Wang2009b} where $a$ denotes the lattice constant of the considered  material. In this work, $q_c = 50/\delta$ is adopted. 
% ------------------------------------------------------------------------
% Calcul Asymptotique : Rousseau
% ------------------------------------------------------------------------
\subsection{Approximate calculation}
\label{sec:Chap5-ExtNF}
Recently, Rousseau et al. \cite{Rousseau2009,Rousseau2010,Rousseau2012} derived, under few simple conditions, an asymptotic expression of the NF RHF $p$-polarized evanescent contribution. This contribution is considered for two reasons. First, it dominates the other contributions in extreme near-field regime for dielectrics and some other materials such as HD-Si for instance. Second, its exact calculation is the most resource-consuming due to the unknown and eventually large cutoff wave vector in some situations.\\
First, they started, when considering a small temperature difference $\delta T$ between the two planes, by defining a radiative NF exchange coefficient $h$ :
\begin{eqnarray}
h & = &  \frac{\dot{q}(\delta,T)}{\delta T}
\label{eq:Chap5-RousseauPropEvanExchangeCoeff}
\end{eqnarray}
which can be written as the sum of two coefficients $h_{prop}$ and $h_{evan}$ corresponding to the propagative and evanescent contributions respectively. Let's focus on the $i$-polarized ($i \in \{p,s\}$) monochromatic evanescent contribution to the radiative transfer exchange coefficient which is given by :
\begin{eqnarray}
h^i_{evan} (\omega) & = & \left[ \int_{\frac{\omega}{c}}^{\infty} \frac{d^2 q}{(2 \pi)^2} 
\overbrace{\left(4 \times e^{2  \imath \gamma_3 \delta} 
\frac{{Im}(r_{31}^i) {Im}(r_{32}^i)}{|1-r_{31}^i r_{32}^i e^{2 \imath \gamma_3 \delta} |^2} \right)}^{\tau_{evan}^i(r^i_{3j},\delta)} \right]
\times \frac{\partial \Theta(\omega,T)}{\partial T}\\
& = & 2 \pi h^0(\omega,T) 
\times \int_{\frac{\omega}{c}}^{\infty} \frac{q dq}{q_0^2} \tau_{evan}^i(r^i_{3j},\delta)
\label{eq:Chap5-RousseauEvanExchangeCoeff}
\end{eqnarray} 
where $\tau_{evan}^i(r^i_{3j},\delta)$ is ($\omega,\mathbf{q}$) mode transmission probability from medium $1$ to medium $2$ \cite{Biehs2010,benabdallah2010} and 
\begin{eqnarray}
h^0 (\omega,T) & = & \frac{q_0^2}{4 \pi^2} \frac{\partial \Theta}{\partial T}
\\
 & = & \frac{1}{T} \frac{\hbar \omega}{k T} \frac{\hbar \omega^3}{4 \pi^2 c^3}
 \left(\frac{1}{2 \; \sinh(\frac{\hbar \omega}{2 k T})} \right) ^2
\label{eq:Chap5-RousseauH0}
\end{eqnarray}
is proportional to the Planck function derivative. 
If we consider the electrostatic regime, i.e. $q \gg q_0= \frac{\omega}{c}$, $p$-polarization Fresnel coefficients become independent of $q$ since they tend toward $r^p = \frac{\epsilon - 1}{\epsilon + 1}$. Then, we can show \cite{Rousseau2012} that $h_{evan}^p$, prevailing in our case, may be written as :
\begin{eqnarray}
h_{evan}^p (u,T,\delta) & = & \frac{3}{2 \pi^3} \frac{g_0}{d^2} h^0(u) \times \frac{{Im}(r^p_{31}) {Im}(r^p_{32})}{{Im}(r^p_{31}r^p_{32})} {Im}(Li_2(r^p_{31}r^p_{32}))
\label{eq:Chap5-Rousseauhspp}
\end{eqnarray}
\nomenclature[L]{$u$}{Reduced frequency}%
where $Li_2$ is the dilogarithm function (see \cite{Abramowitz1965} for definition and \cite{Osacar1995} for numerical evaluation), $h^0(u) = \frac{u^2}{(e^u-1)^2}$, $u = \frac{\hbar \omega}{k T}$ and $g_0 = \frac{\pi^2 k^2 T}{3 h}$ is the quantum of heat conduction.
\nomenclature[L]{$g_0$}{Quantum of heat conduction}%
\\
NF heat flux is then given by :
\begin{eqnarray}
\dot{q}(T,\delta) & \simeq & \left( \int_0^{\infty} h_{evan}^p (u,T,\delta) du \right) \times \delta T
\label{eq:Chap5-RousseauAsymptoticFlux}
\end{eqnarray}
Therefore, the heat flux calculation is reduced to a simple integral evaluation and the problem of the cutoff wave vector is apparently resolved. Given the assumed hypotheses in order to obtain expressions \ref{eq:Chap5-Rousseauhspp} and \ref{eq:Chap5-RousseauAsymptoticFlux}, a verification with an exact calculation of results obtained by this method might be necessary.
% ------------------------------------------------------------------------
% Resultats
% ------------------------------------------------------------------------
\section{Optimization State of the art}
Different groups have already tackled the question of maximizing the NF radiative heat transfer, for plane-plane configuration in particular. Zhuoming Zhang's group of Georgia Tech. has been particularly prolific.
First, Basu et al. \cite{Basu2008} led a theoretical parametric study of radiative transfer between two semi-infinite planes of HD-Si. This material was considered because of its interesting optical properties that can be controlled through the doping level \cite{Basu2008,Marquier2004b,Nefzaoui2012}. In fact, its dielectric permittivity is modeled by a Drude model where the doping concentration controls both of the plasma frequency $\omega_p$ and the damping coefficient $\Gamma$. 
\nomenclature[G]{$\Gamma$}{Damping factor}%
\nomenclature[G]{$\omega_p$}{Plasma frequency}%
They observed that the RHF spectrum presents a peak around the plasma frequency and a blue-shift of the peak position when the doping concentration increases. They also noted that the total exchanged RHF increases with doping until a maximum that depends on temperature and $\omega_p$. At room temperature, this optimum is observed for a doping concentration between $10^{19}$ and $10^{20}$ (cm$^{-3}$). Let us note that similar results, obtained by a different approach, have been reported for HD-Si by \cite{Rousseau2009}. Finally, they considered two planes with different doping concentrations and tend towards the conclusion that the maximal RHF is obtained for identical media. 
Then, in another work \cite{Basu2009}, they went beyond the particular case of HD-Si by considering two identical semi-infinite planes of a completely fictive material. They found that the dielectric permittivity maximizing the exchanged RHF can be written $\epsilon=-1 + \imath \delta$ with  ${Im}{(\epsilon)} = \delta \ll 1$. It is worth noting here that this form of $\epsilon$ underlies a hypothesis of a non-dispersive medium.
At the same time, Wang et al. \cite{Wang2009b} considered less restrictive situations and generalized first results previously obtained for HD-Si to other real materials (SiC, MgO) and fictive materials modeled by Drude and Lorentz models. For Drude model, control parameters are $\omega_p$ and $\Gamma$ and the high frequency limit of the dielectric permittivity $\epsilon_{\infty}$. 
\nomenclature[S]{$\infty$}{High frequency limit}%
Lorentz model has an additional parameter $\omega_0$ which corresponds to the frequency of transverse optical phonons. 
\nomenclature[G]{$\omega_0$}{Transverse optical phonons circular frequency (also usually referred to by $\omega_{TO}$)}%
Authors make the following general conclusions : (1) Drude model leads to higher values of maximal RHF than Lorentz model. For this reason, Lorentz model presents its highest performances when $\omega_0 = 0$, i.e. when it is equivalent to Drude model. That is why we focus on Drude model in the following points. (2) For Drude model : (2-1) Lower values of $\epsilon_{\infty}$ lead to the highest values of maximal RHF. These values are the closest to the condition given by \cite{Basu2009} and previously presented. (2-2) At room temperature, a maximum of RHF is observed for $\omega_p \simeq 10^{14}$ (rad.s$^{-1}$) and $\Gamma / \omega_p \simeq 0,1$. The position of this maximum is strongly $T$-dependent. In addition, the maximum is realized by a compromise between the peak width (controlled by $\Gamma$) and the peak position (controlled by $\omega_p$).\\
More recently, several authors exploited graphene features to enhance NF RHF. Graphene presents palsmon-polaritons in the terahertz domain which makes it particularly interesting for radiative heat transfer around room temperature\cite{Svetovoy2012}. Besides, more than HD-Si, its optical properties can be tuned with doping level or chemical potential. Finally, graphene dielectric function is non-local, i.e. its dielectric permittivity in general, and its plasma frequency in particular, depend on the wave vector. Therefore, it presents a big variety of resonant modes which may allow to consider their coupling with other materials resonant modes.    
These authors showed that a thin film of graphene deposited on a dielectric that does not support surface phonon-polaritons leads to an enhancement of the exchanged NF RHF between two semi-infinite planes of the same graphene-covered material by three and almost four orders of magnitude. However, this enhancement decreases rapidly with temperature and is spectacular only for temperatures lower than room temperature. 
Another group from l'Institut d'Optique of Paris \cite{Messina2012b}, showed for a plane-plane system of SiC, that a thin film of graphene on the surface of one of the two planes leads to additional peaks in the spectrum of the local density of states due to the coupling of graphene modes with those of SiC. These modes contribute to the increase of exchanged NF RHF.\\
Shall we here emphasize practical potential of graphene as a selective emitter for NF TPV devices. Indeed, the possibility to tune graphene plasmon-polariton resonance frequencies would allow their adjustment to the band gap of different photovoltaic converters. Messina et al. actually demonstrated \cite{Messina2012a} for a TPV device composed of a boron nitride emitter (at $T_e = 450$ K) and an indium antimonide cell, that a graphene film with a chemical potential of $0.5$ eV on the surface of the cell leads to an increase of the maximal efficiency of the system by a factor two to reach $\eta \simeq 20 \%$ and an increase of output power by almost one order of magnitude. Higher performances corresponding to higher operating temperatures in the range [$600,1200$] K have been recently presented by another group of the MIT \cite{Ilic2012} who considered a slightly different system where graphene plays the role of a selective emitter. Prior works had already considered NF TPV devices based on metallic selective emitters such as tungsten \cite{Laroche2006,Francoeur2011,Basu2009bRev} but graphene seems to monopolize the community recent attention due to the diversity of potential applications it makes possible thanks to the "flexibility" of its surface modes and optical properties.
%***************************************************************************************************
% Résultats
%***************************************************************************************************
\section{Results}
Formalisms presented in the first section are used to calculate the exchanged NF RHF between two semi-infinite planes separated by a distance $\delta = 10$ nm.
Planes dielectric functions are modeled by local Drude and Lorentz models usually adopted to describe real materials (gap thickness considered here is much larger than non-local phenomena onset distances \cite{Chapuis2008,Ezzahri2013}). Calculations are made for both identical and different planes around $300$ K while varying  models parameters in their usual variation ranges with three main goals in mind : (1) For identical planes : to determine optical properties, fictive in this case, that would maximize NF RHF in order to guide, for a given application, the choice of a real material to use or the design of meta-materials (2) For different planes : to verify the hypothesis which states that the maximal RHF is obtained when the two planes materials are identical (3) To compare the accuracy and the resource-consumption cost of the exact and approximate methods.
%***************************************************************************************************
% Modèle de Drude
%***************************************************************************************************
\subsection{Drude model}
\label{sec:Chap5-Drude}
We remind the expression of the dielectric permittivity in this model :
\begin{eqnarray}
\epsilon({\omega}) & = & \epsilon_{\infty} - \frac{\omega_p^2}{\omega^2+ \imath \Gamma \omega}
\label{eq:Chap5-DrudeModel}
\end{eqnarray}
where $\epsilon_{\infty}$ is the high frequency limit of the dielectric permittivity, $\omega_p$ the plasma frequency and $\Gamma$ the damping coefficient.
%***************************************************************************************************
% Modèle de Drude : Single Material
%***************************************************************************************************
\subsubsection{Identical media}
\begin{figure}[ht]
\begin{center}
        \begin{subfigure}[b]{0.47\textwidth}
                \centering
                \includegraphics[width=\textwidth]{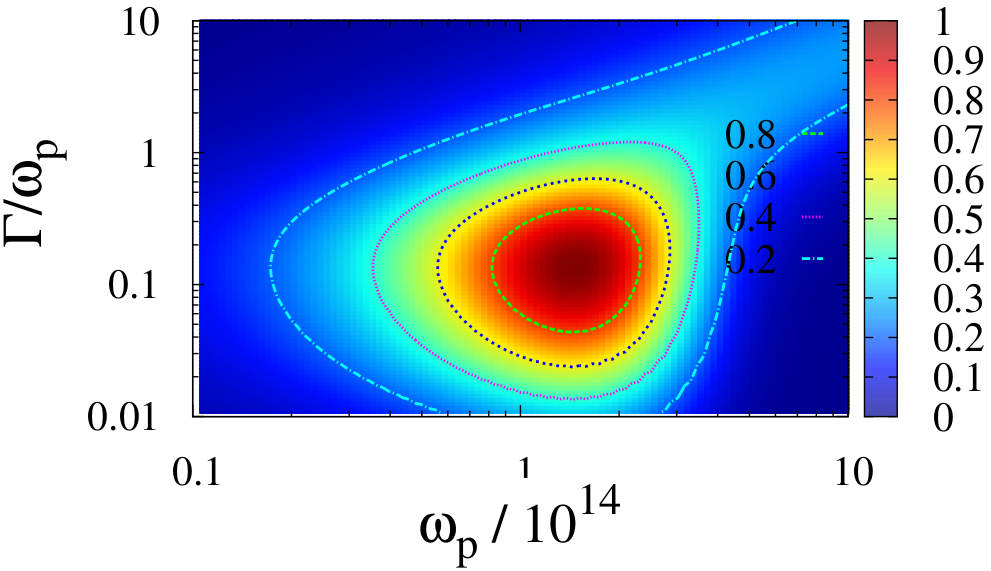}
                % OMp * 10^{-14} , Gam/OMp , Phi_max
                %   1.4454396596536672       0.13803842191347615        220425.39162241269
%                \caption{$\epsilon_{\infty} = 1$, $\dot{q}_{max} = 220425$ (W.m$^{-2}$) pour $\omega_p \times 10^{-14} = 1,44$ et $\frac{\gamma}{\omega_p} = 0,138 $.}
				\caption{}
                \label{fig:Chap5-FullDrudeFluxEpsInf1}
        \end{subfigure} 
        \quad
        \begin{subfigure}[b]{0.47\textwidth}
                \centering
                \includegraphics[width=\textwidth]{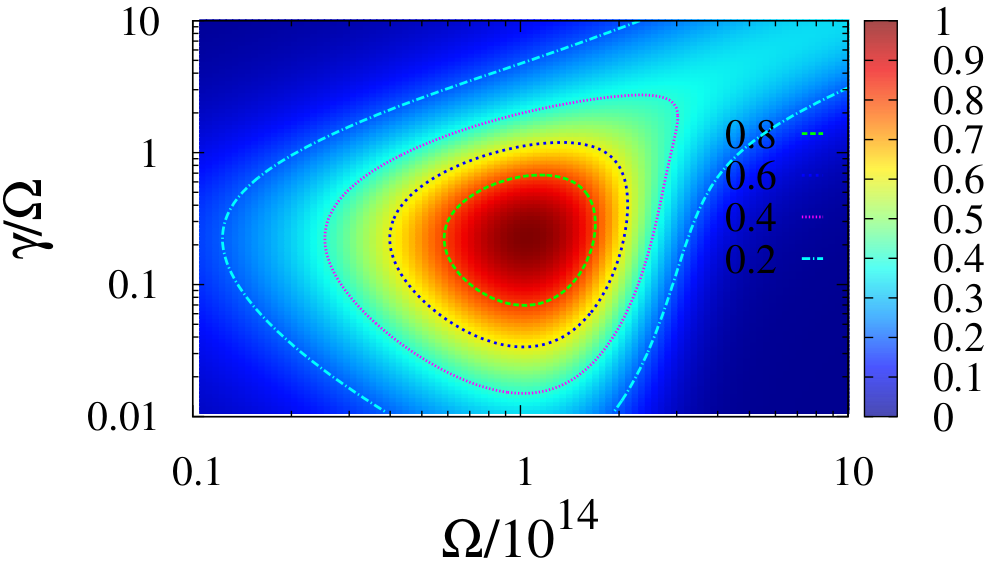}
				% OMp * 10^{-14} , Gam/OMp , Phi_max
				%    1.0471285020627683       0.23988330572949451        229208.35600676507                
%                \caption{$\epsilon_{\infty} = 1$, $\dot{q}_{max} = 229208$ (W.m$^{-2}$) pour $\omega_p \times 10^{-14} = 1,05$ et $\frac{\gamma}{\omega_p} = 0,24 $}
				\caption{}
                \label{fig:Chap5-AsympDrudeUmesh1000-EpsInf1}
        \end{subfigure}
\caption{Normalized NF RHF between two identical semi-infinite planes. The dielectric permittivity is given by Drude model with $\epsilon_{\infty} = 1$. Results are obtained by both exact calculation (a) and asymptotic one (b).}
\label{fig:Chap5-DrudeSingle}
\end{center}
\end{figure}

\begin{figure}[ht]
\begin{center}
\centering
\includegraphics[width=0.6\textwidth]{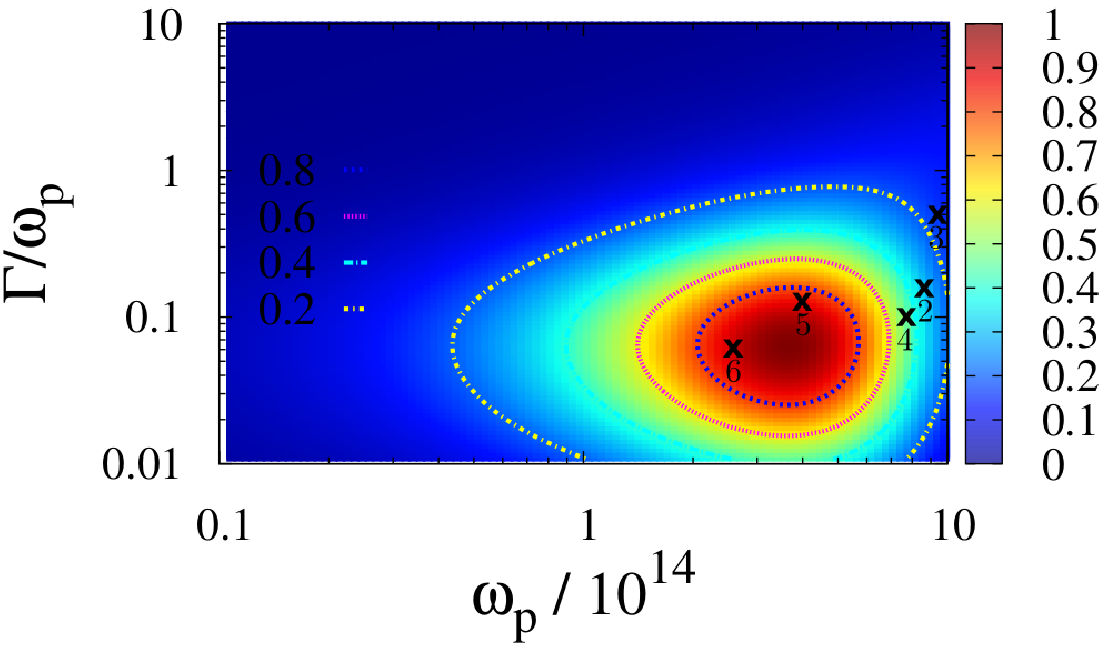}
\caption{Normalized NF RHF for $\epsilon_{\infty} = \epsilon_{\infty,Si} = 11.8$ obtained by exact calculations. Crosses represent $p$ and $n$-type HD-Si at different doping concentrations (see Table \ref{tab:Chap5-DopedSiUsualDrudeParams} for parameters values of the different points).}
\label{fig:Chap5-FFDrudeSingleEpsInf118DSi}
\end{center}
\end{figure}

~~\\
First, we calculate NF RHF between two identical semi-infinite planes modeled by Drude model with $\epsilon_{\infty} = 1$. $\omega_p$ and $\Gamma$ are varied within the ranges $[10^{13},10^{15}]$ and $[10^{-2} \times \omega_p, 10 \times \omega_p]$ respectively which cover these parameters ranges for HD-Si. Some values of these parameters for HD-Si with doping concentration around $10^{19}$ cm$^{-3}$ are given in Table \ref{tab:Chap5-DopedSiUsualDrudeParams}. Media $1$ and $2$ are considered at $300$ K and  $299$ K respectively.
\begin{table}[h!]
\begin{center}
\begin{small}
\begin{tabular}{cccccc}
\hline
N{$^{\circ}$} & Doping type & Concentration $\times 10^{-19}$($cm^{-3}$) & $\epsilon_{\infty}$ & $\omega_{p} \times 10^{-14}$
(rad.s$^{-1}$) & $\frac{\Gamma}{\omega_p}$   \\
\hline 
1 & Si:B & $27$ & $11.8$ & $16$ & $6.7 \times 10^{-1}$ \\ 
2 & Si:B & $6.7$ & $11.8$ & $8.3$ & $1.7 \times 10^{-1}$ \\
3 & Si:P & $10$ & $11.8$ & $9.7$ & $5.1 \times 10^{-1}$ \\ 
4 & Si:P & $5.3$ & $11.8$ & $7.28$ & $10^{-1}$ \\
5 & Si:P & $1.6$ & $11.8$ & $4$ & $1.3 \times 10^{-1}$ \\
6 & Si:P & $0.52$ & $11.8$ & $2.3$ & $6 \times 10^{-2}$ \\
\hline
\end{tabular}
\end{small}
\end{center} 
\caption{Drude model parameters for the dielectric permittivity of $p$ and $n$-type HD-Si with bore (Si:B) and phosphorus (Si:P) respectively \cite{Borghesi1985}.}
\label{tab:Chap5-DopedSiUsualDrudeParams}
\end{table}
\nomenclature[L]{$n$}{Mesh points number}%
%***************************************************************************************************
\paragraph{Plasma frequency and damping effects} 
~~\\
We present in figure \ref{fig:Chap5-DrudeSingle} the normalized RHF exchanged between the two media. Plotted results are obtained by both exact calculation (Figure \ref{fig:Chap5-FullDrudeFluxEpsInf1}) and asymptotic calculation in the case of extreme NF with the electrostatic limit approximation (Figure \ref{fig:Chap5-AsympDrudeUmesh1000-EpsInf1}). Only the dominating $p$-polarization is presented here.
First, we can note the actual existence of a maximum (See Table \ref{tab:Chap5-DrudeOptimalParameters} for its value and coordinates.).
\begin{table}[h!]
\begin{center}
\begin{small}
\begin{tabular}{cccccccc}
\hline Method & $\epsilon_{\infty}$ & $\omega_p \times 10^{-14}$ (s$^{-1}$) & $\frac{\Gamma}{\omega_p}$ & $\dot{q}_{max}$ (W.m$^{-2}$) & $n_{\omega,q}$ & $n_{\omega_p,\Gamma}$ & $t$ (s) \\ 
\hline
E & $1$ & $1.51$ & $0.17$ & $229336$ & $500$ & $100$ & $32921$ \\ 
A & $1$ & $1.05$ & $0.24$ & $229208$ & $1000$ & $100$ & $23$\\
%   2.5118862936122772       3.46736867584146596E-002   78648.425461718216 
\hline E & $5$ & $2.51$ & $3.7 \times 10^{-2}$ & $78656$ & $500$ & $100$ & $14864$\\ 
A & $5$ & $0.79$ & $0.11$ & $78676$ & $1000$ & $100$ & $13$ \\
\hline E & $10$ & $3.47$ & $1.51 \times 10^{-2}$ & $42123$ & $400$ & $100$  & $13952$  \\ 
A & $10$ & $0.76$ & $6.91 \times 10^{-2}$ & $43128$ & $1000$& $100$ & $13$\\
\hline E & $20$ & $4.57$ & $2.29 \times 10^{-3}$ & $24269$ & $400$ & $100$ & $16961$\\ 
A & $20$ & $0.72$ & $3.71 \times 10^{-2}$ & $22621$ & $1000$ & $100$ & $13$\\
\hline
\end{tabular}
\end{small}
\end{center}
\caption{Drude model parameters maximizing the exchanged NF RHF between two semi-infinite planes separated by a gap of thickness $\delta = 10$ nm for different values of $\epsilon_{\infty}$. Values are obtained by both exact (E) and approximate (A) calculations. $n_{\omega,q}$ and $n_{\omega_p,\Gamma}$ are mesh points numbers for ($\omega$,$q$) modes and for control parameters ($\omega_p$ and $\Gamma$) respectively. $t$ is CPU calculation time to obtain the corresponding figures.}
\label{tab:Chap5-DrudeOptimalParameters}
\end{table}
Beyond the maximum position, these figures reveal the RHF sensitivity to the different parameters.
In fact, we observe that a relative variation between $22 \%$ and $27 \%$ for $\omega_p$ and of about  $50 \%$ for $\Gamma$ gives values of the flux larger than $0.95 \times \dot{q}_{max}$. These parameters values admissible variations to keep high flux values are slightly larger than those reported in literature\cite{Basu2010a}.
\begin{figure}[ht]
\begin{center}
        \begin{subfigure}[b]{0.45\textwidth}
                \centering
                \includegraphics[width=\textwidth]{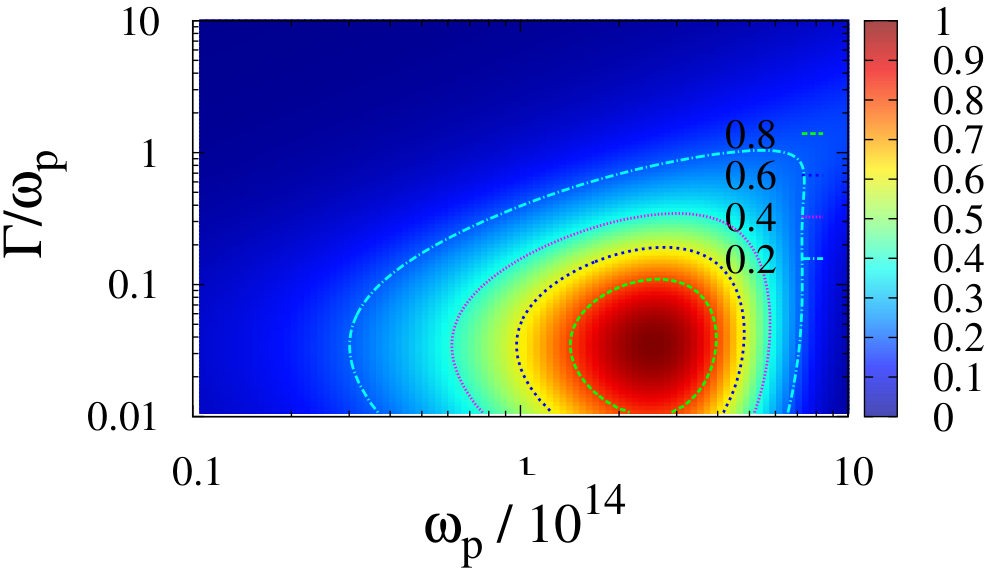}
                % OMp * 10^{-14} , Gam/OMp , Phi_max
                %    2.5118862936122772       3.71535222978220739E-002   78656.109425203729                  
%                \caption{$\epsilon_{\infty} = 5$, $\dot{q}_{max} = 78656$ (W.m$^{-2}$) pour $\omega_p \times 10^{-14} = 2,51$ et $\frac{\gamma}{\omega_p} \times 10^{2} = 3,71$.}
				\caption{}
                \label{fig:Chap5-FullDrudeFluxEpsInf5}
        \end{subfigure}
        \quad
        \begin{subfigure}[b]{0.45\textwidth}
                \centering
                \includegraphics[width=\textwidth]{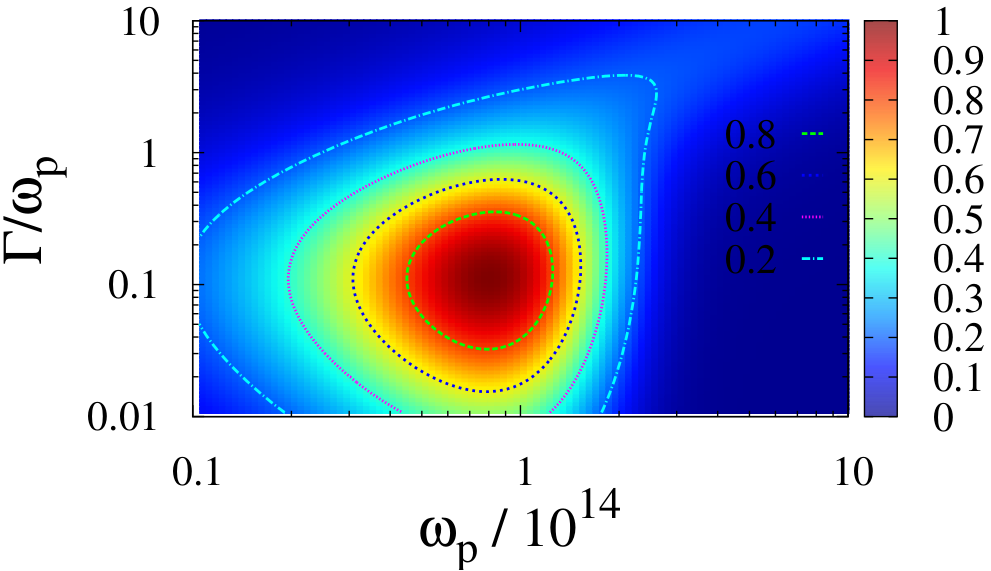}
				% OMp * 10^{-14} , Gam/OMp , Phi_max
				%  0.79432819111732544       0.11220184081046163        78675.868683920999                
%                \caption{$\epsilon_{\infty} = 5$, $\dot{q}_{max} = 78676$ (W.m$^{-2}$) pour $\omega_p \times 10^{-14} = 0.79$ et $\frac{\gamma}{\omega_p} = 0,11 $.}
				\caption{}
                \label{fig:Chap5-AsympDrudeUmesh1000-EpsInf5}
        \end{subfigure}

        \begin{subfigure}[b]{0.45\textwidth}
                \centering
                \includegraphics[width=\textwidth]{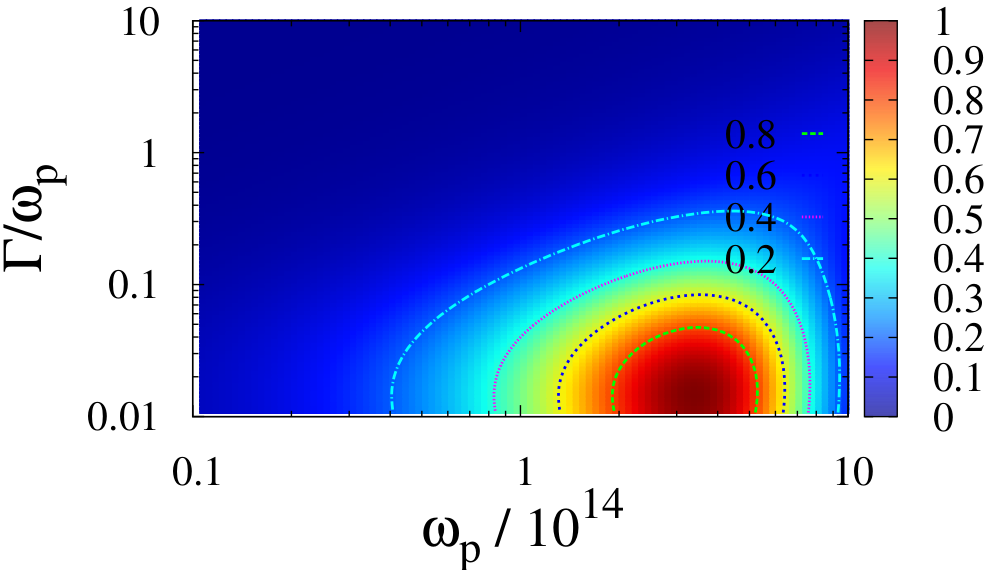}
                % OMp * 10^{-14} , Gam/OMp , Phi_max
                %    3.4673681999632939       1.51356123441454410E-002   43123.961921059134
%                \caption{$\epsilon_{\infty} = 10$, $\dot{q}_{max} = 43124$ (W.m$^{-2}$) pour $\omega_p \times 10^{-14} = 3,47$ et $\frac{\gamma}{\omega_p} \times 10^{2} = 1,51 $.}
				\caption{}
                \label{fig:Chap5-FullDrudeFluxEpsInf10}
        \end{subfigure} 
        \quad
        \begin{subfigure}[b]{0.45\textwidth}
                \centering
                \includegraphics[width=\textwidth]{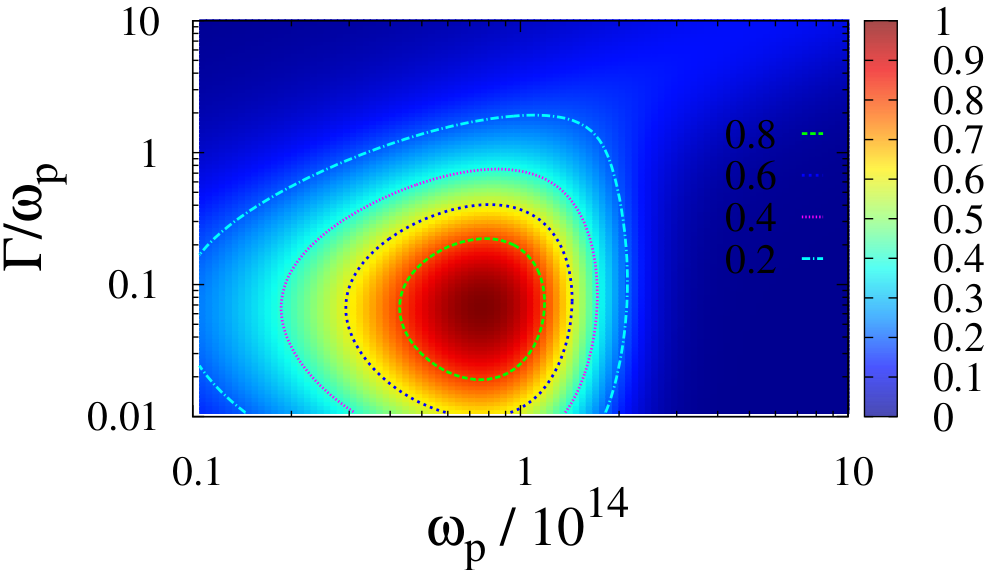}
                % OMp * 10^{-14} , Gam/OMp , Phi_max
                %   0.75857756670031973       6.91830976615947363E-002   43128.883621286404
%                \caption{$\epsilon_{\infty} = 10$, $\dot{q}_{max} = 43129$ (W.m$^{-2}$) pour $\omega_p \times 10^{-14} = 0.76$ et $\frac{\gamma}{\omega_p} \times 10^2 = 6,9 $.}
				\caption{}
                \label{fig:Chap5-AsympDrudeUmesh1000-EpsInf10}
        \end{subfigure}   
\caption{Normalized NF RHF between two semi-infinite planes modeled by Drude model for $\epsilon_{\infty} = 5$ : (a - b) and $\epsilon_{\infty} = 10$ : (c - d) obtained by exact (left column) and approximate (right column) calculations.}
\label{fig:Chap5-DrudeSingleEpsInfVar}
\end{center}
\end{figure}
%%%%%%%%%%%%%%%%%%%%%%%%%%%%%%%%%%%%%%%%%%%%%%%%%%%%%%%%%%%%%%%%%%%%%%%%%%%%%%%%%%%%%%%%%%%%%%%%%%%%
\paragraph{High frequency limit of the dielectric permittivity effect}
~~\\ 
Similar calculation results are presented in Figure \ref{fig:Chap5-DrudeSingleEpsInfVar} for $\epsilon_{\infty} = 5$ (Figures \ref{fig:Chap5-AsympDrudeUmesh1000-EpsInf5} and \ref{fig:Chap5-FullDrudeFluxEpsInf5}) and $\epsilon_{\infty} = 10$ (figures \ref{fig:Chap5-AsympDrudeUmesh1000-EpsInf10} and \ref{fig:Chap5-FullDrudeFluxEpsInf10}). We observe as Basu et al.\cite{Basu2010a}, a decrease of the maximal flux value when $\epsilon_{\infty}$ increases. In fact, lower values of $\epsilon_{\infty}$ lead to lower values of ${Re} (\epsilon) = \epsilon_{\infty} - {Re}({\frac{\omega_p^2}{\omega^2+ \imath \Gamma \omega}})$ which are the closet to fit Basu et al. condition to maximize the NF RHF\cite{Basu2009}, i.e. ${Re}(\epsilon) = -1$.
%%%%%%%%%%%%%%%%%%%%%%%%%%%%%%%%%%%%%%%%%%%%%%%%%%%%%%%%%%%%%%%%%%%%%%%%%%%%%%%%%%%%%%%%%%%%%%%%%%%%
\paragraph{Exact versus approximate calculation}
~~\\
Maximal values of the RHF obtained by both methods are almost the same with a relative error around $10^{-4}$ (see Table \ref{tab:Chap5-DrudeOptimalParameters}).
In the case $\epsilon_{\infty} = 1$, maxima are realized for $(\omega_p,\frac{\Gamma}{\omega_p}) = (1.51 \times 10^{14} \; rad.s^{-1}, 1.7 \times 10^{-1})$ and $(\omega_p,\frac{\Gamma}{\omega_p}) = (1.05 \times 10^{14} \; rad.s^{-1}, 0.24 \times 10^{-1})$ with exact and approximate calculations respectively. The relative error on positions is quite important, up to $27 \%$ and $82 \%$ for $\omega_p$ and $\Gamma$ respectively. An exact calculation of the flux value corresponding to approximate optimal parameters is $30 \%$ lower than the actual maximal flux value. This discrepancy on optimal parameters given by both methods increases with $\epsilon_{\infty}$.
Let us note however the resource-consumption gain made by the use of the asymptotic approximation : figure \ref{fig:Chap5-FullDrudeFluxEpsInf1} (exact) was obtained in $32921$ (s) versus $23$ (s) for figure \ref{fig:Chap5-AsympDrudeUmesh1000-EpsInf1} (approximate), i.e. a ratio of almost $1500$ between the two. This ratio particularly depends on the parallel wave vector mesh resolution and increases rapidly with it. Calculations were made on an Intel\textsuperscript{\textregistered} Xeon\textsuperscript{\textregistered} E5620 @ 2.40GHz, with 12288 Kb of cache and 4 Go of RAM memory.
\begin{figure}[ht]
\begin{center}
        \begin{subfigure}[b]{0.47\textwidth}
                \centering
                \includegraphics[width=\textwidth]{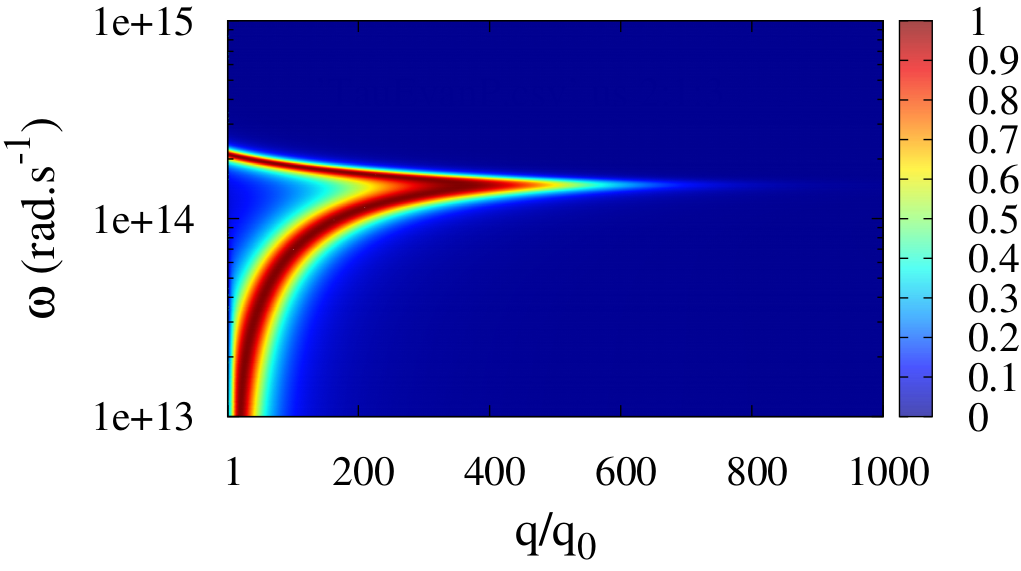}
				\caption{}
                \label{fig:Chap5-Dispersion-EpsInf1-DrudeExactOptTau}
        \end{subfigure} 
        \quad
        \begin{subfigure}[b]{0.47\textwidth}
                \centering
                \includegraphics[width=\textwidth]{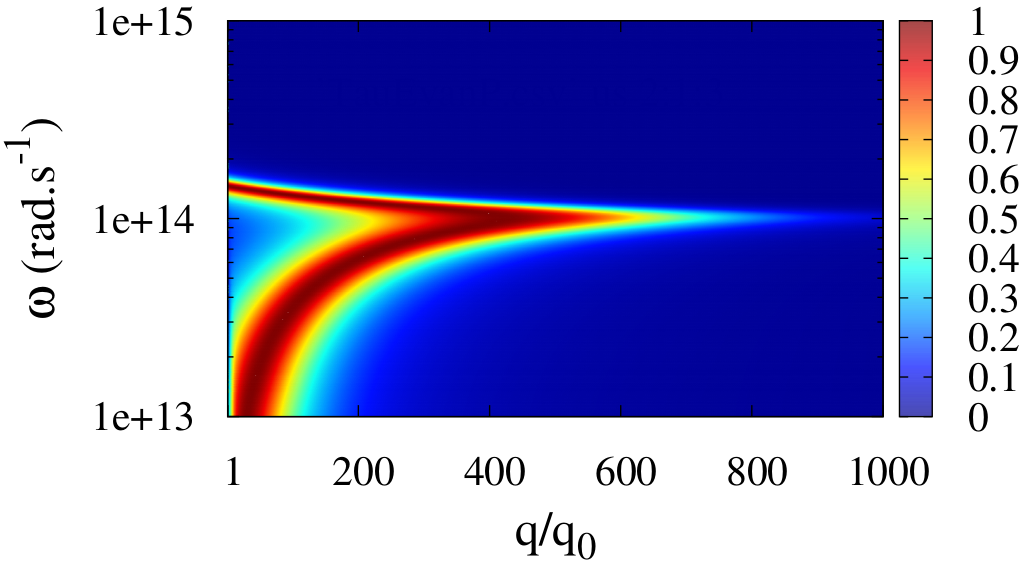}
				\caption{}
                \label{fig:Chap5-Dispersion-EpsInf1-DrudeApproxOptTau}
        \end{subfigure}
\caption{Transmission probability of $p$-polarized evanescent modes $\tau_{evan}^p(\omega,\mathbf{q})$ as defined in equation \ref{eq:Chap5-RousseauEvanExchangeCoeff} for Drude model with $\epsilon_{\infty} = 1$. Two cases are considered : (a) the exact optimum given in (Table \ref{tab:Chap5-DrudeOptimalParameters}, line $1$) and (b) the approximate one given in (Table \ref{tab:Chap5-DrudeOptimalParameters}, line $2$).}
\label{fig:Chap5-DrudeDispersionTau}
\end{center}
\end{figure}\\ 
In order to understand the origin of the discrepancy between the two methods, we plot in figure \ref{fig:Chap5-DrudeDispersionTau} each $p$-polarized evanescent mode $(\omega,\mathbf{q})$ transmission coefficient $\tau_{evan}^p(\omega,\mathbf{q})$ as defined in equation \ref{eq:Chap5-RousseauEvanExchangeCoeff}. We consider the exact optimum (Table \ref{tab:Chap5-DrudeOptimalParameters}, line $1$, Figure \ref{fig:Chap5-Dispersion-EpsInf1-DrudeExactOptTau}) and the approximate one (Table \ref{tab:Chap5-DrudeOptimalParameters}, line $2$, Figure \ref{fig:Chap5-Dispersion-EpsInf1-DrudeApproxOptTau}). Several observations can be made : (1) The approximate optimum presents a high transmission coefficient (red and yellow areas) for more numerous modes than the exact one. (2) This higher number of transmitted modes is more pronounced for modes with a large wave vector parallel component $q$. (3) The exact optimum presents less transmitted modes but at a higher circular frequency.\\
According to these observations, it is obvious that the discrepancy between the optimum position given by both methods is due to the fact that the electrostatic approximation ignores modes with low $q$. The optimum position shift in the approximate approach also induces a decrease in each mode mean energy. This decrease is compensated in the overall flux density by a larger modes number. In order to accurately estimate each mode contribution, the value of the integrand $(q / q_0) \times \tau_{evan}^p (\omega,q)$ of the sum over $q$ which appears in equation \ref{eq:Chap5-RousseauEvanExchangeCoeff} is more relevant than the mere modes transmission probability. It is plotted in figure \ref{fig:Chap5-DrudeDispersionTauKappa}.
\begin{figure}[ht]
\begin{center}
        \begin{subfigure}[b]{0.47\textwidth}
                \centering
                \includegraphics[width=\textwidth]{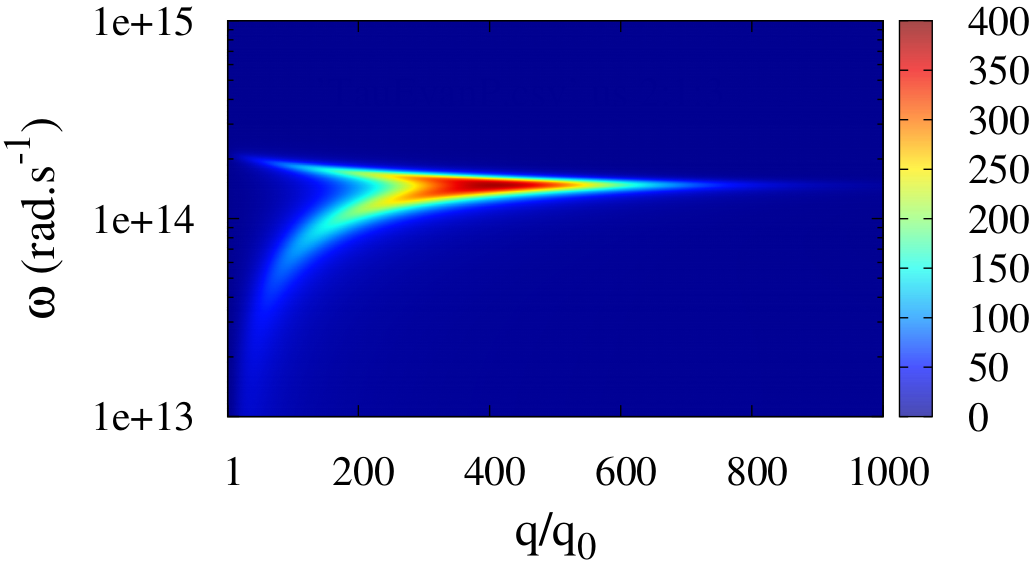}
				\caption{}
                \label{fig:Chap5-Dispersion-EpsInf1-DrudeExactOptTauKappa}
        \end{subfigure} 
        \quad
        \begin{subfigure}[b]{0.47\textwidth}
                \centering
                \includegraphics[width=\textwidth]{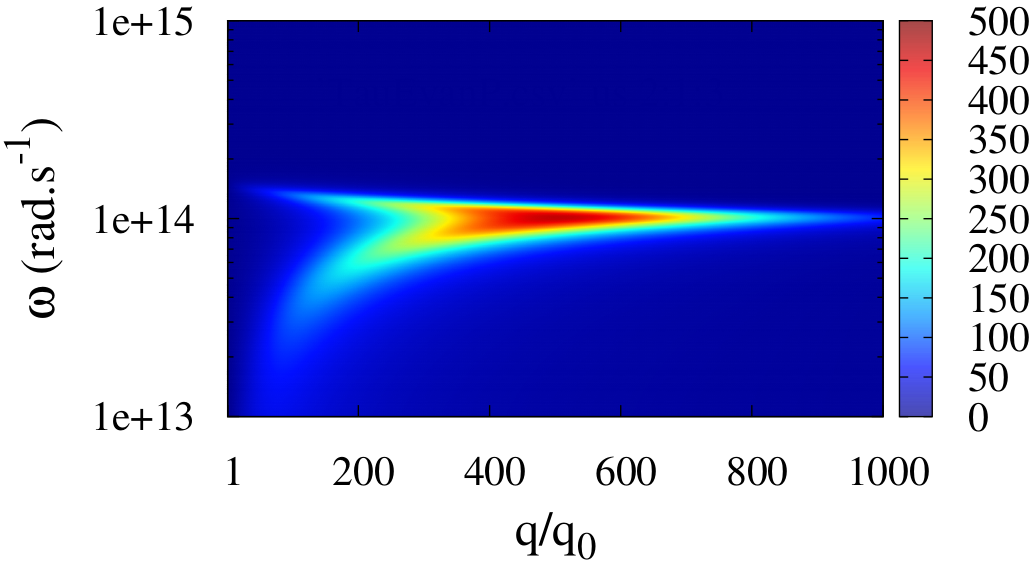}
				\caption{}
                \label{fig:Chap5-Dispersion-EpsInf1-DrudeApproxOptTauKappa}
        \end{subfigure}
\caption{The integrand $(q / q_0) \times \tau_{evan}^p (\omega,q)$ of the sum over the wave vector parallel component in equation \ref{eq:Chap5-RousseauEvanExchangeCoeff} for Drude model with $\epsilon_{\infty} = 1$. The same cases as in figure \ref{fig:Chap5-DrudeDispersionTau} are considered : (a) the exact optimum given in (table \ref{tab:Chap5-DrudeOptimalParameters}, line $1$) and (b) the approximate one given in (table \ref{tab:Chap5-DrudeOptimalParameters}, line $2$).}
\label{fig:Chap5-DrudeDispersionTauKappa}
\end{center}
\end{figure}\\
It appears through this figure that the weight of high wave number modes is dominating. These modes, for both exact and approximate optima, lay in the same $q / q_0$ range,  $q / q_0 \in [200,600]$ in this case, but for slightly different circular frequencies however which may explain the small relative error on flux density values obtained by both methods. Finally, figure  \ref{fig:Chap5-DrudeDispersionTauKappa} allows an accurate calculus of the cutoff wave vector value. If we define $q_c$ as the largest wave number verifying $(q_c / q_0) \times \tau_{evan}^p (\omega_p,q_c) = \frac{1}{2} \left[ (q / q_0) \times \tau_{evan}^p (\omega_p,q) \right]_{max}$, we obtain $(q_c / q_0) = 597$ and $(q_c / q_0) = 768$ which leads to $q_c \simeq 3 / \delta$ and $q_c \simeq 2.7 / \delta$ for the exact and the approximate calculation respectively. The cutoff wave vector is hence of the order of $1 / \delta$ and was actually overestimated in our first calculations.
%%%%%%%%%%%%%%%%%%%%%%%%%%%%%%%%%%%%%%%%%%%%%%%%%%%%%%%%%%%%%%%%%%%%%%%%%%%%%%%%%%%%%%%%%%%%%%%%%%%%
\paragraph{Case of heavily doped silicon at $300$ K}
~~\\ 
Figure \ref{fig:Chap5-FFDrudeSingleEpsInf118DSi} presents similar results for $\epsilon_{\infty} = \epsilon_{\infty,Si} = 11.8$. The aim here is to determine whether HD-Si, previously considered by several authors\cite{Basu2010a,Rousseau2009} to maximize NF RHF, is well adapted to this task. For this reason, parameters values corresponding to HD-Si are represented on the same figure by crosses. Previously reported results \cite{Basu2010a,Rousseau2009} stating a maximal flux for a doping concentration between $10^{19}$ and $10^{20}$ (cm$^{-3}$) are more likely to be confirmed. Besides, we can state according to this figure that HD-Si around $10^{19}$ (cm$^{-3}$) is a good candidate to NF RHF maximization at room temperature since it allows to reach almost $0.9 \times \dot{q}_{max}$ that can be obtained with a Drude model with $\epsilon_{\infty} = 11.8$ (we obviously assume that $\epsilon_{\infty}$ is a parameter that can hardly be varied).
%***************************************************************************************************
% Modèle de Drude : Double Material
%***************************************************************************************************
\subsubsection{Non-identical media}
~~\\
\begin{figure}[ht]
\begin{center}
\centering
\includegraphics[width=0.6\textwidth]{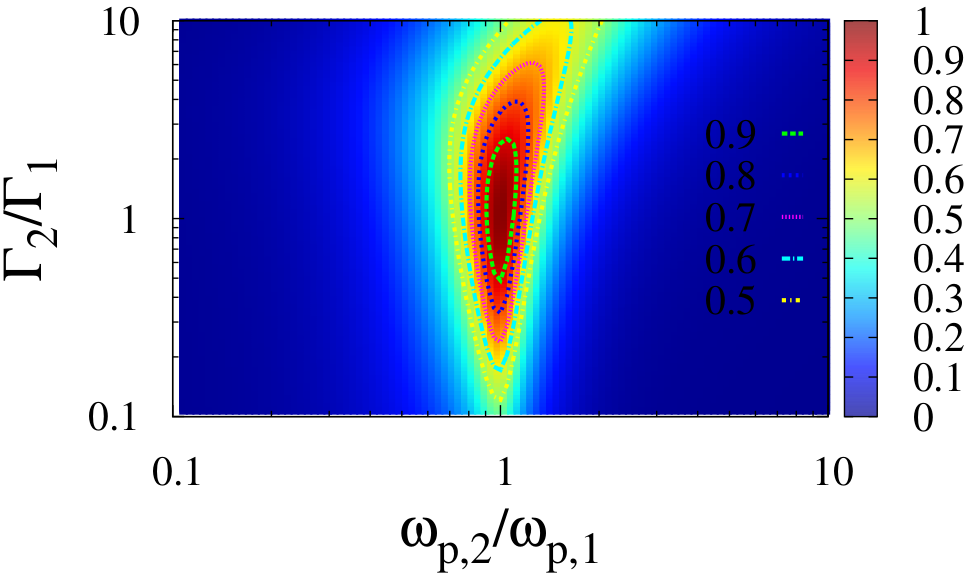}
\caption{Normalized NF RHF exchanged by two semi-infinite planes of different optical properties modeled by Drude model as function of plasma frequencies ratio and damping factors ratio obtained by exact calculation. The figure is obtained by exact calculations.}
\label{fig:Chap5-2MatDrude}
\end{center}
\end{figure}
~~\\
The only change considered in this paragraph lies in the fact that exchanging semi-infinite planes dielectric functions are not identical while they are still modeled by a Drude model. We are aiming to a double objective : (1) verify the statement of maximal flux for identical media due to a more efficient coupling of identical modes supported by the same materials (2) See in what extent, a more or less important difference in optical properties of considered materials affects the exchanged NF RHF. This second objective has an obvious applied interest since real materials eventually used in a particular application are never exactly identical.
For this sake, we consider the two media around $300$ K separated by $\delta=10$ nm.
We also consider, without generality loss, $\epsilon_{\infty} = 1$. Medium $1$ parameters are fixed to optimal values previously obtained (Table \ref{tab:Chap5-DrudeOptimalParameters}, line $1$). Control parameters are then the second medium Drude model parameters, i.e. $\omega_{p,2}$ and $\Gamma_2$. 
Figure \ref{fig:Chap5-2MatDrude} presents the normalized exchanged NF RHF as a function of plasma frequencies ratio $\omega_{p,2} / \omega_{p,1}$ and damping factors ratio $\Gamma_2 / \Gamma_1$.\\
The maximum is actually realized for $(\omega_{p,2} / \omega_{p,1},\Gamma_2 / \Gamma_1) = (1,1)$, i.e. for identical media. Besides, the flux value is more sensitive to $\omega_p$ than to $\Gamma$ value. In fact, the flux is maintained at high values ($\dot{q} > 0.9 \times \dot{q}_{max}$) for $\omega_{p,2} / \omega_{p,1} \in [0.9,1.1]$ and $\Gamma_2/\Gamma_1 \in [0.49,2.53]$. A $10 \%$ variation of $\omega_{p}$ leads to a comparable variation of the flux value. The same flux variation is obtained with a variation of $\Gamma$ up to $150 \%$. However, the asymmetry of $\dot{q}$-behavior as a function of $\Gamma$ is worth noting. In fact, the sign of $\Gamma$-variation affects strongly the variation of the flux. Finally, the flux sensitivity to $\omega_p$ decreases for larger values of $\Gamma$. This is due to the fact that $\Gamma$ controls the exchanged flux spectral density peak width \cite{Basu2010a} : the larger $\Gamma$ the larger the peak width which allows looser constraints on the peak position controlled by $\omega_p$.
%***************************************************************************************************
% Modèle de Lorentz
%***************************************************************************************************
\subsection{Lorentz model}
First, we remind the dielectric permittivity expression according to this model\cite{Palik1985} :
\begin{eqnarray}
\epsilon(\omega) & = & \epsilon_{\infty} - \frac{\omega_p^2}{\omega^2+ \imath \Gamma \omega - \omega_0^2}
\label{eq:Chap5-LorentzModel}
\end{eqnarray}
where $\omega_p^2 = \omega_{LO}^2 - \omega_0^2$ with $\omega_{LO}$ the longitudinal optical phonons circular frequency, $\omega_0 = \omega_{TO}$ the transverse optical phonons circular frequency and $\Gamma$ the damping factor.\\
\nomenclature[G]{$\omega_{LO}$}{Longitudinal optical phonons circular frequency}%
%%%%%%%%%%%%%%%%%%%%%%%%%%%%%%%%%%%%%%%%%%%%%%%%%%%%%%%%%%%%%%%%%%%%%%%%%%%%%%%%%%%%%%%%%%%%%%%%%%%%
\subsubsection{Identical media}
~~\\
First, identical media are considered, medium $1$ at $300$ K and medium $2$ at $299$ K. The gap thickness between the two planes is $\delta = 10$ nm. Compared to Drude model, Lorentz model has an additional parameter, transverse optical phonons frequency $\omega_0$ in this case. In this study, $\omega_0 = \omega_{0,SiC} = 1.49 \times 10^{14}$ (rad.s$^{-1}$) \cite{Palik1985} is considered constant which reduces the problem to a two-parameter problem. Control parameters are $\omega_{LO}$ and $\Gamma$. Results will be presented as a function of $\frac{\omega_{LO}}{\omega_0}$ and $\frac{\Gamma}{\omega_0}$. 
%%%%%%%%%%%%%%%%%%%%%%%%%%%%%%%%%%%%%%%%%%%%%%%%%%%%%%%%%%%%%%%%%%%%%%%%%%%%%%%%%%%%%%%%%%%%%%%%%%%%
\paragraph{Longitudinal phonons frequency and damping factor effect}
\begin{figure}[ht]
\begin{center}
        \begin{subfigure}[b]{0.47\textwidth}
                \centering
                \includegraphics[width=\textwidth]{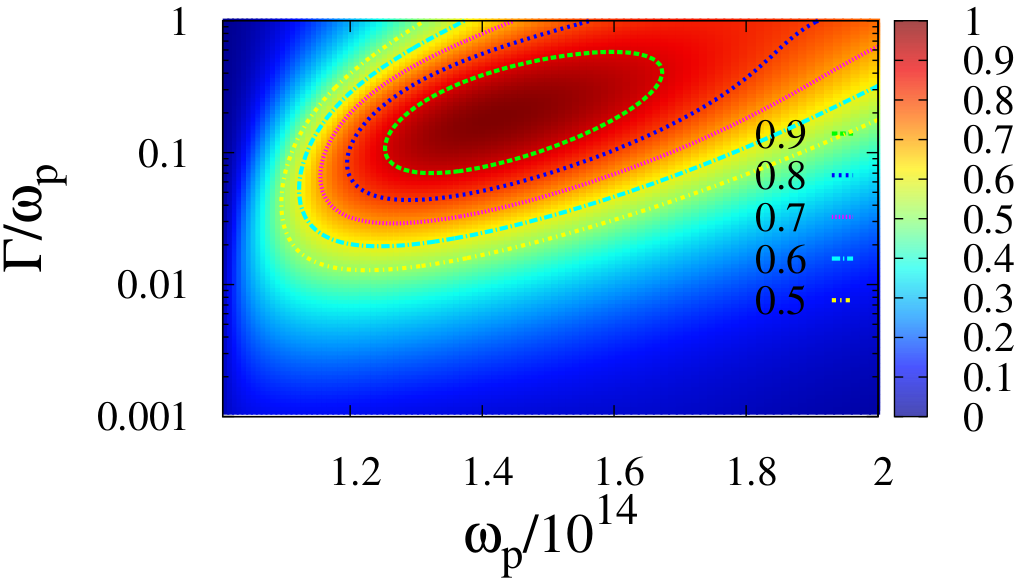}
                \caption{}
                \label{fig:Chap5-FFSingleLorentz-EpsInf1}
        \end{subfigure} 
		\quad        
        \begin{subfigure}[b]{0.47\textwidth}
                \centering
                \includegraphics[width=\textwidth]{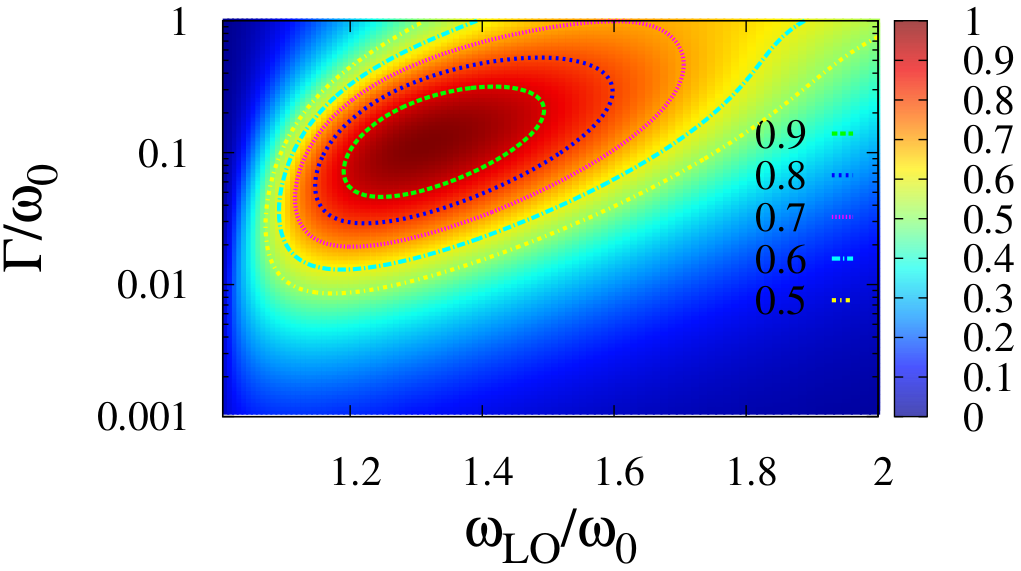}
                \caption{}
                \label{fig:Chap5-AsympSingleLorentz-EpsInf1}
        \end{subfigure}   
                 
        \begin{subfigure}[b]{0.47\textwidth}
                \centering
                \includegraphics[width=\textwidth]{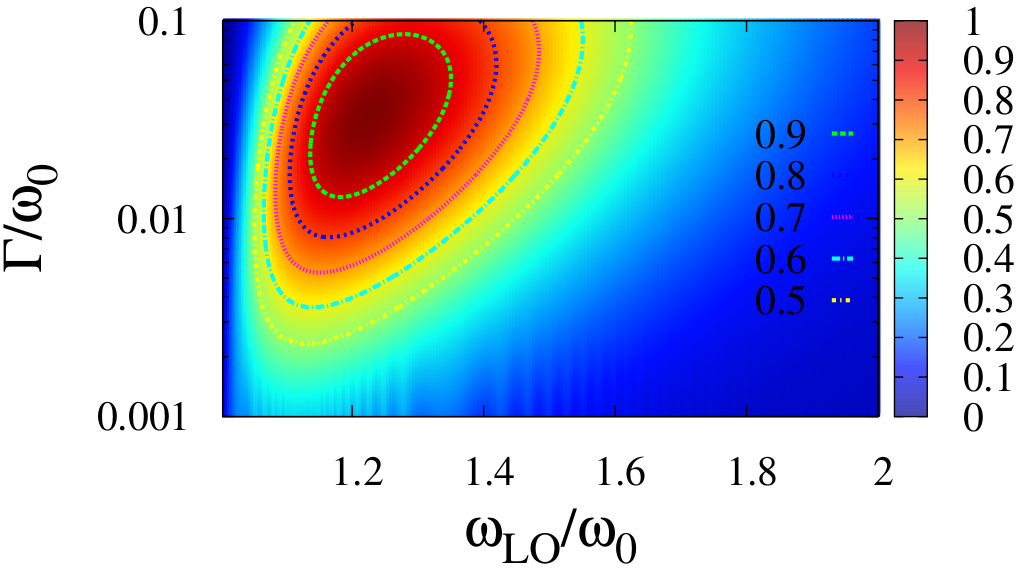}
                \caption{}
                \label{fig:Chap5-FFSingleLorentz-EpsInf10}
        \end{subfigure}   
        \quad
       \begin{subfigure}[b]{0.47\textwidth}
                \centering
                \includegraphics[width=\textwidth]{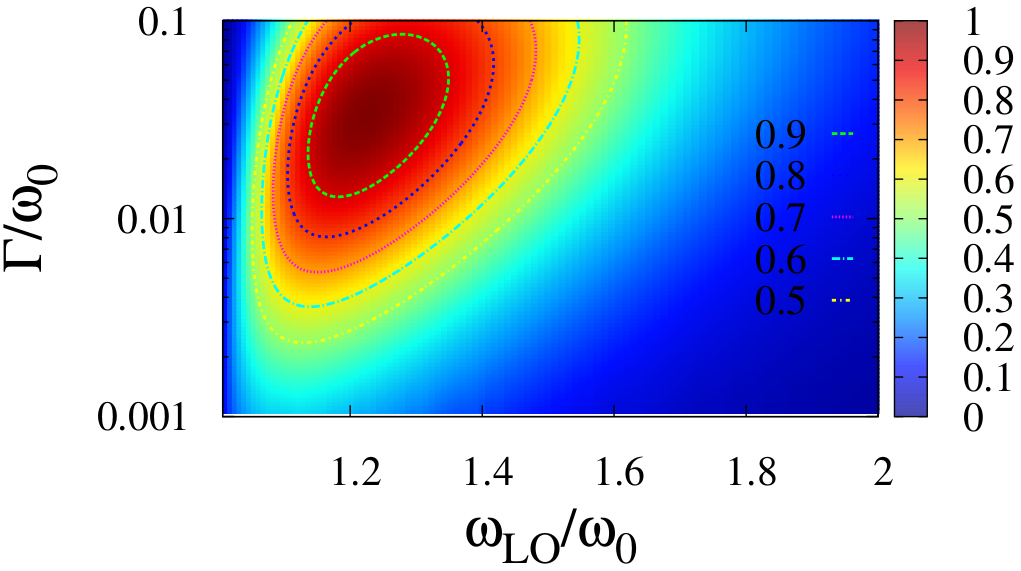}
                \caption{}
                \label{fig:Chap5-AsympSingleLorentz-EpsInf10}
        \end{subfigure}  
\caption{Normalized NF RHF for $\epsilon_{\infty} = 1$ (a-b) and $\epsilon_{\infty} = 10$ (c-d) obtained by exact (left column) and approximate (right column) calculations. See Table \ref{tab:Chap5-LorentzSingleOptimalParameters} for a summary of principal relevant results.}
\label{fig:Chap5-SingleLorentzEpsInfVar}
\end{center}
\end{figure}
~~\\
Figure \ref{fig:Chap5-SingleLorentzEpsInfVar} presents the normalized NF RHF exchanged by two semi-infinite planes which dielectric permittivities are modeled by Lorentz model for 
$\epsilon_{\infty} = 1$ (\ref{fig:Chap5-FFSingleLorentz-EpsInf1}-\ref{fig:Chap5-AsympSingleLorentz-EpsInf1}) and $\epsilon_{\infty} = 10$ (\ref{fig:Chap5-FFSingleLorentz-EpsInf10}-\ref{fig:Chap5-AsympSingleLorentz-EpsInf10}) obtained by exact (left column) and asymptotic calculations (right column).
Principal relevant results of this figure, concerning the maximum position and value as well as calculation time, are summarized in Table \ref{tab:Chap5-LorentzSingleOptimalParameters}.\\
As for Drude model, we observe the existence of a maximum which is realized by a compromise between the phonons frequencies and the damping factor, i.e. between the peak position and width.
For $\epsilon_{\infty} = 1$ for instance, a maximal flux density $\dot{q}_{max} = 54529$ W.m$^{-2}$ is observed at $(\frac{\omega_{LO}}{\omega_{LO}} = 1.42, \frac{\Gamma}{\omega_{TO}} = 1.9 \times 10^{-1})$. Let us note that this $\dot{q}_{max}$ value is almost five times lower than the value obtained with a Drude model at $\epsilon_{\infty} = 1$ ($\dot{q}_{max} = 229336$ W.m$^{-2}$). In fact, Drude model is the Lorentz model limit when $\omega_0$ goes zero. Thus, we observe, even though the detailed study of this parameter is not presented in the present paper, an increase of the maximal achievable flux with a Lorentz model when $\omega_0$ decreases.
Furthermore, we observe that $\dot{q}_{max}$ sensitivity to $\omega_{LO}$ is much larger than to $\Gamma$. In fact, the flux is kept at relatively high values ($\dot{q} > 0.9 \times \dot{q}_{max}$) with a relative variation of $\omega_{LO}$ around $\pm 12 \%$ ($\omega_{LO} / \omega_{TO} \in [1.25,1.67]$) versus a relative variation of $\Gamma$ up to $+200 \%$ ($\Gamma / \omega_{TO} \in [7\times10^{-2},5.9\times10^{-1}]$).
\begin{table}[h!]
\begin{center}
\begin{small}
\begin{tabular}{cccccccc}
\hline Method & $\epsilon_{\infty}$ & $\frac{\omega_{LO}}{\omega_{TO}}$ & $\frac{\Gamma}{\omega_{TO}}$ & $\dot{q}_{max}$ (W.m$^{-2}$) & $n_{\omega,q}$ & $n_{\omega_p,\Gamma}$ & $t$ (s) \\ 
%   1.4240501861835779       0.19054605924362855        56905.195884983150        1.0000000000000000 
\hline E & $1$ & $1.42$ & $1.9 \times 10^{-1}$ & $56896$ & $2000$ & $100$ & $2.8 \times 10^5$ \\ 
A & $1$ & $1.42$ & $1.9 \times 10^{-1}$ & $56905$ & $10000$ & $100$ & $2.51 \times 10^2$\\
\hline E & $\epsilon_{\infty,SiC} = 6.7$ & $1.24$ & $4.78 \times 10^{-2}$ & $14874$ & $4000$ & $100$ & $1.09 \times 10^6$\\ 
A & $\epsilon_{\infty,SiC} = 6.7$ & $1.24$ & $4.78 \times 10^{-2}$ & $14849$ & $10000$& $100$ & $1.25 \times 10^{2}$ \\
\hline E & $10$ & $1.22$ & $3.31 \times 10^{-2}$ & $10415$ & $3000$ & $100$ & $6.17 \times 10^5$\\ 
A & $10$ & $1.22$ & $3.31 \times 10^{-2}$ & $10391$ & $10000$ & $100$ & $2.5 \times 10^2$\\
\hline
\end{tabular}
\end{small}
\end{center}
\caption{Optimal Lorentz model parameters for identical media plane-plane configuration. Values are obtained by exact (E) and approximate (A) calculations. $\omega_0 = \omega_{0,SiC} = 1.49 \times 10^{14}  \; rad.s^{-1}$ is kept constant. $t$ is CPU time, $n_{\omega,q}$ and $n_{\omega_p,\Gamma}$ are the mesh points number for frequency and wave vector ($\omega$ and $q$)  and control parameters ($\Gamma / \omega_{0}$ and $\omega_{OL} / \omega_{0}$) discretization respectively.}
\label{tab:Chap5-LorentzSingleOptimalParameters}
\end{table}
%%%%%%%%%%%%%%%%%%%%%%%%%%%%%%%%%%%%%%%%%%%%%%%%%%%%%%%%%%%%%%%%%%%%%%%%%%%%%%%%%%%%%%%%%%%%%%%%%%%%
\paragraph{High frequency limit of the dielectric function effect}
~~\\
As for Drude model and for the same reasons, we observe a decrease of $\dot{q}_{max}$ when $\epsilon_{\infty}$ increases in addition to a shift of the maximum position to lower values of $\omega_{LO} / \omega_{0}$ and $\Gamma / \omega_{0}$.
%%%%%%%%%%%%%%%%%%%%%%%%%%%%%%%%%%%%%%%%%%%%%%%%%%%%%%%%%%%%%%%%%%%%%%%%%%%%%%%%%%%%%%%%%%%%%%%%%%%%
\paragraph{Exact versus approximate calculation}
~~\\
At this point, Lorentz model strongly contrasts with what was previously observed with Drude model giving very accurate asymptotic results for the maximal flux value as well as for its position.
The position relative error is lower than $10^{-3}$. Similarly low relative error values are observed for the maximal flux value, except for the case $\epsilon_{\infty} = 10$ where it reaches $2.3 \times 10^{-3}$. Indeed, the maximal flux relative error increases with $\epsilon_{\infty}$, i.e. when $\dot{q}_{max}$ decreases. A part of this error is due to the omission of the propagative contribution in the asymptotic calculation. This contribution is almost constant for different values of $\epsilon_{\infty}$ while the $p$-polarized evanescent contribution and the total flux decrease when $\epsilon_{\infty}$ increases. \\
In spite of comparable accuracy, asymptotic calculations are still $1000$ times faster than exact ones.
In addition, the approximate method shows a better convergence. In fact, some numerical oscillations due to slow convergence can be observed on Figure \ref{fig:Chap5-FFSingleLorentz-EpsInf10} for small flux values.
\begin{figure}[ht]
\begin{center}
        \begin{subfigure}[b]{0.47\textwidth}
                \centering
                \includegraphics[width=\textwidth]{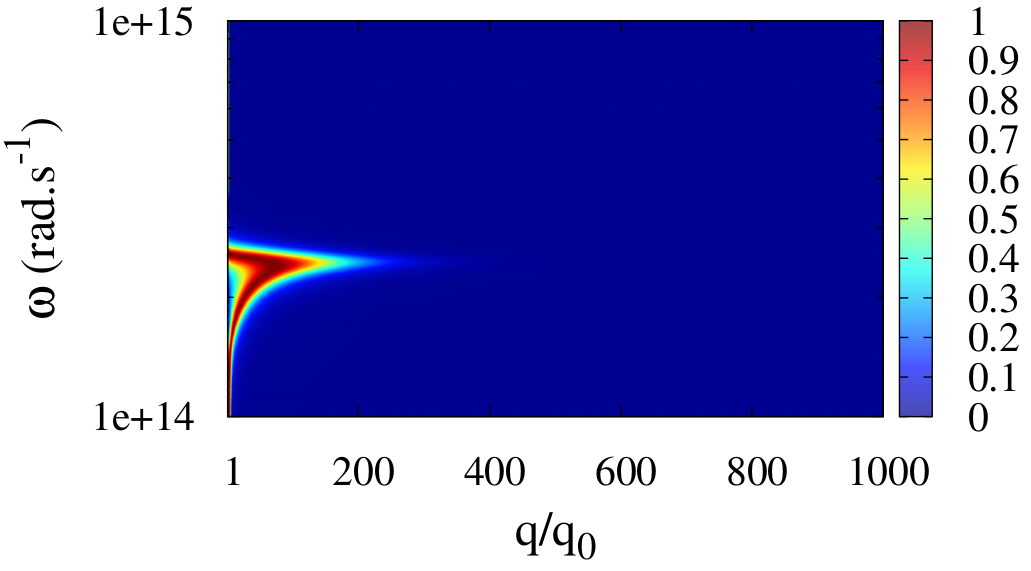}
				\caption{}
                \label{fig:Chap5-Dispersion-EpsInf1-LorentzOptTau}
        \end{subfigure} 
        \quad
        \begin{subfigure}[b]{0.47\textwidth}
                \centering
                \includegraphics[width=\textwidth]{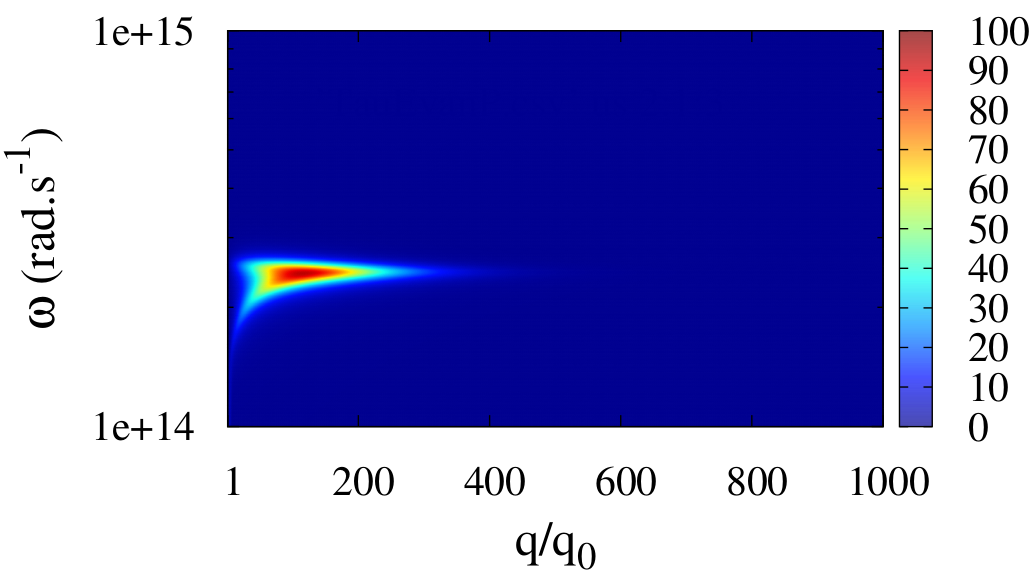}
				\caption{}
                \label{fig:Chap5-Dispersion-EpsInf1-LorentzOptTauKappa}
        \end{subfigure}
\caption{(a) Transmission probability of $p$-polarized evanescent modes $\tau_{evan}^p(\omega,q)$ and (b) the integrand $(q / q_0) \times \tau_{evan}^p (\omega,q)$ of the sum over the wave vector parallel component in equation \ref{eq:Chap5-RousseauEvanExchangeCoeff} for Lorentz model with $\epsilon_{\infty} = 1$. Lorentz model parameters are given in (Table \ref{tab:Chap5-LorentzSingleOptimalParameters}, line $1$). Only one case, the exact optimum in this case, is considered here since approximate calculations gave similar results with high accuracy.}
\label{fig:Chap5-LorentzDispersion}
\end{center}
\end{figure}\\
To understand the origin of the consistency of the two methods in the case of Lorentz model we proceed as done previously for Drude model and examine the transmission probability $\tau_{evan}^p(\omega,q)$ and the integrand $(q / q_0) \times \tau_{evan}^p (\omega,q)$ of the sum over the wave vector parallel component in equation \ref{eq:Chap5-RousseauEvanExchangeCoeff}. These two quantities are plotted in figures \ref{fig:Chap5-Dispersion-EpsInf1-LorentzOptTau} and \ref{fig:Chap5-Dispersion-EpsInf1-LorentzOptTauKappa} respectively. We first note that, compared to Drude model, modes are transmitted here in a much lower number which explains the lower flux density values. Second, transmitted modes mainly lay in the range $q / q_0 \in [15,150]$ if we consider $\tau_{evan}^p(\omega,q) \geq \frac{1}{2}$. This concentration of transmitted modes around medium and high $q$ values is behind the high accuracy of the approximate method. Finally, the cutoff wave vector, considering the same criterion as in Drude model, is found for $q_c/q_0 = 216$, i.e. for $q_c \simeq 1.8 / \delta$ which is of the order of $1 / \delta $.
%%%%%%%%%%%%%%%%%%%%%%%%%%%%%%%%%%%%%%%%%%%%%%%%%%%%%%%%%%%%%%%%%%%%%%%%%%%%%%%%%%%%%%%%%%%%%%%%%%%%
\paragraph{Case of silicon carbide at $300$ K}
~~\\
Finally, we consider the case of silicon carbide (SiC). This material has been extensively studied in NF radiative heat transfer literature for its strong surface phonon-polariton resonances around $\omega = 10^{14}$ (rad.s$^{-1}$).
\begin{figure}[ht]
	\begin{center}
	\includegraphics[width=0.6\textwidth]{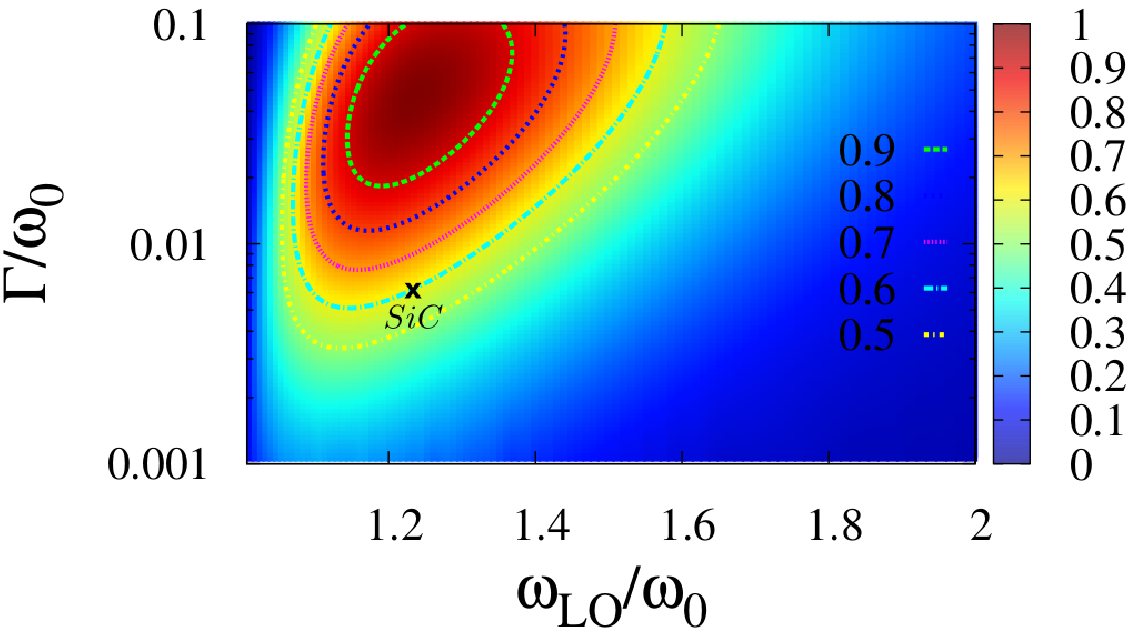}
	\caption{Normalized NF RHF between two semi-infinite planes as a function of $\frac{\omega_{LO}}	{\omega_{0}}$ and $\frac{\Gamma}{\omega_{0}}$ for $\epsilon_{\infty} = \epsilon_{\infty,SiC} = 6.7$.
SiC parameters values correspond to the point ($1.24,6 \times 10^{-3}$)\cite{Palik1985} indicated by a cross.}
	\label{fig:Chap5-SingleLorentzEpsInfSiC}
	\end{center}
\end{figure}        
~~\\
Figure \ref{fig:Chap5-SingleLorentzEpsInfSiC} presents the normalized NF RHF exchanged by two semi-infinite planes modeled by Lorentz model with $\epsilon_{\infty} = \epsilon_{\infty,SiC} = 6.7$. 
With only $0.6 \times \dot{q}_{max}$, SiC is far from approaching Lorentz model optimal performances unlike the case of HD-Si which parameters allowed flux values as high as $90\%$ of the maximal RHF that can be obtained with a Drude model when $\epsilon_{\infty} = \epsilon_{Si}$.
Besides, this figure is obtained by calculations with higher resolution meshes, $n_{\omega,q} = 4000$ in this case. Compared to figure \ref{fig:Chap5-FFSingleLorentz-EpsInf10}, this shows that numerical oscillations magnitude decreases slowly when the mesh points number $n_{\omega,q}$ of ($\omega,\mathbf{q}$) space increases. CPU time is however one order of magnitude larger than previously, i.e. than in figure \ref{fig:Chap5-FFSingleLorentz-EpsInf10}.
%%%%%%%%%%%%%%%%%%%%%%%%%%%%%%%%%%%%%%%%%%%%%%%%%%%%%%%%%%%%%%%%%%%%%%%%%%%%%%%%%%%%%%%%%%%%%%%%%%%%
\subsubsection{Non-identical media}
\begin{figure}[ht]
\begin{center}
        \begin{subfigure}[b]{0.47\textwidth}
                \centering
                \includegraphics[width=\textwidth]{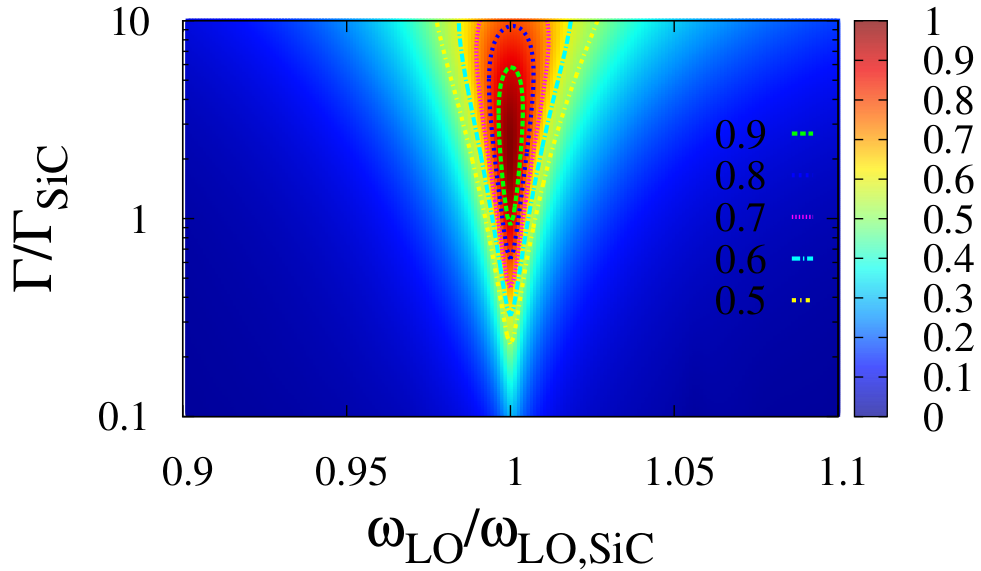}
                \caption{}
                \label{fig:Chap5-Asymp2MatLorentzFullFluxSiCX}
        \end{subfigure}     
        \quad
        \begin{subfigure}[b]{0.47\textwidth}
                \centering
                \includegraphics[width=\textwidth]{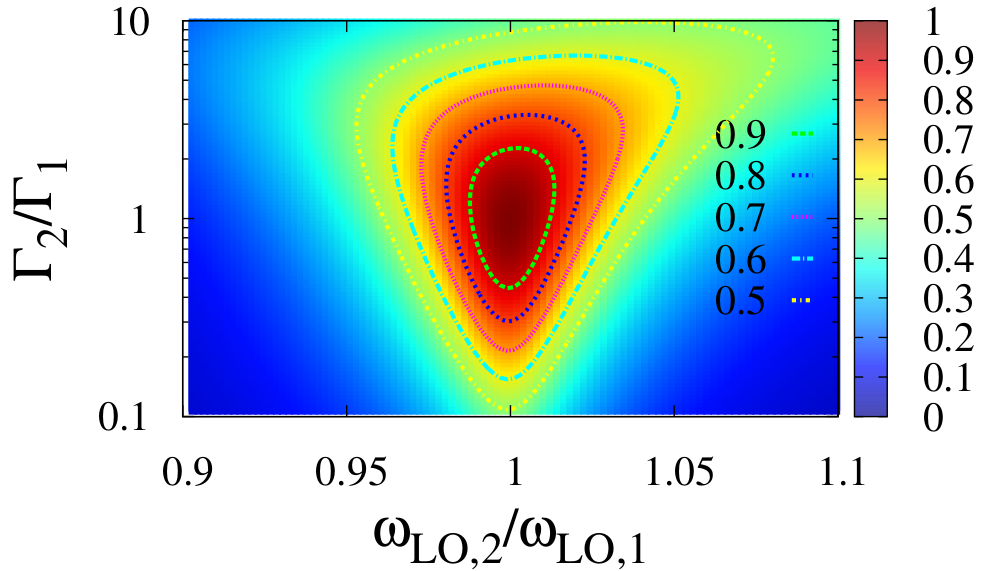}
                \caption{}
                \label{fig:Chap5-Asymp2MatLorentzFullFluxEpsInfSiCX}
        \end{subfigure}     
\caption{Normalized NF RHF exchanged by two semi-infinite planes of different non-identical materials which dielectric functions are modeled by Lorentz models. Two cases are considered : (a) SiC exchanging with another material, (b) the material maximizing the transfer between two identical media at $\epsilon_{\infty} = 6.7$ (see Table \ref{tab:Chap5-LorentzSingleOptimalParameters}, line 3) with another material.}
\label{fig:Chap5-2MatLorentz}
\end{center}
\end{figure}

~~\\
Now, consider two semi-infinite planes made of non identical materials. We will analyze two cases : (1) the case of SiC and a slightly different material (Figure \ref{fig:Chap5-Asymp2MatLorentzFullFluxSiCX}) (2) The case of the fictive material realizing the optimal performances with $\epsilon_{\infty} = \epsilon_{\infty,SiC} = 6.7$ (see Table \ref{tab:Chap5-LorentzSingleOptimalParameters}, line 3) that we will note material $1$ with a slightly different material (Figure \ref{fig:Chap5-Asymp2MatLorentzFullFluxEpsInfSiCX}). For both cases, $\omega_{0} = \omega_{TO,SiC} = 1.49 \times 10^{14} \; rad.s^{-1}$ is constant. Control parameters are then $\omega_{OL} / \omega_{OL,SiC}$ and $\Gamma / \Gamma_{SiC}$ in the first case and $\omega_{OL} / \omega_{OL,1}$ and $\Gamma / \Gamma_{1}$ in the second.
~~\\
Unsurprisingly, the optimum is observed at $(1,1)$ in both cases, i.e. for identical media. We also observe a high sensitivity, more pronounced for SiC, of the flux to $\omega_{LO}$. In fact, a relative variation of the order of $10^{-4}$ of $\omega_{LO}$ around the point $(1,1)$ decreases the flux below $0.9 \times \dot{q}_{max}$ while a $8 \times 10^{-3}$ relative variation of $\omega_{LO}$ halves the flux value. $\dot{q}$ is however much less sensitive to $\Gamma$ since relative variations of this parameter in the range [$-8\%,581\%$] maintains $\dot{q} > 0.9 \times \dot{q}_{max}$.
The high asymmetry of this range around zero is due to the fact that $\Gamma$ controls the imaginary part of the dielectric permittivity peak width and height. On the other hand, ${Im}(\epsilon)$ controls both emission and absorption. Thus, if $\Gamma$ increases for the material exchanging with SiC, this material dielectric permittivity imaginary part peak will be wider and lower than SiC's. Therefore, all SiC modes will contribute to the transfer (the new peak is wider than SiC peak), with a lower modes density though (the new peak is lower than SiC peak). Then again, when $\Gamma$ decreases, the peak becomes narrower and higher than SiC's. All SiC modes do not contribute to the transfer anymore while contributing modes have the same density than in SiC-SiC system.\\
Similar considerations can be made about he second case (Figure \ref{fig:Chap5-Asymp2MatLorentzFullFluxEpsInfSiCX}) with the only difference that material $1$ damping factor is lower than SiC's. This implies a wider peak for ${Im}(\epsilon_1)$ which allows looser constraints on the peak position controlled by $\omega_{LO}$ and $\omega_0$. For instance, a $\pm 1.2 \%$ relative variation of $\omega_{LO}$ keeps the flux higher than $0.9 \times \dot{q}_{max}$. This value is two orders of magnitude higher than SiC's, even though it is still relatively small and restrictive in regard to the quality of materials that can be obtained with usual nano-materials deposition techniques.
%%%%%%%%%%%%%%%%%%%%%%%%%%%%%%%%%%%%%%%%%%%%%%%%%%%%%%%%%%%%%%%%%%%%%%%%%%%%%%%%%%%%%%%%%%%%%%%%%%%%
\section{Conclusion}
In this work, a study of the effects of different parameters of usual materials local dielectric functions models (Drude and Lorentz) on NF RHF exchanged by two semi-infinite planes separated by a nanometric gap at room temperature is presented. For this purpose, exact and approximate (according to the asymptotic electrostatic limit approximation in the extreme near-field regime presented in \cite{Rousseau2009}) calculations of the heat flux were calculated. We then showed that the asymptotic approximation leads to highly accurate results, in particular for Lorentz model, with a calculation time at least one thousand times shorter than exact calculation time. Two particular materials usually considered for near-field heat transfer optimization were also considered : silicon carbide (SiC) and highly doped silicon (HD-Si). HD-Si reveals to be well adapted to this aim. In fact, it allows to reach $90 \%$ of maximal achievable heat flux by a Drude model with $\epsilon_{\infty} = 11.8$. It is however possible to overcome these performances by a metamaterial that would have a much lower value of the dielectric permittivity high frequency limit. On the other hand, Lorentz model in general, and SiC in particular, are not the best choice in order to maximize NF radiative heat transfer, at room temperature at least. In addition, SiC is particularly penalizing since its maximal performance is strongly dependent on the quality of used materials. Thus, very small discrepancies, of the order of $10^{-3}$, between the phonons frequencies of the two SiC samples would halve the maximal achievable radiative heat flux.
We also showed, for both models, that the maximal RHF is obtained when the two semi-infinite planes are made of identical materials.\\
Finally, it is worth mentioning the mesoscopic description of NF radiative heat transfer recently developed \cite{benabdallah2010,Biehs2010} and which renews the understanding of this kind of transfer : radiative energy is transported through different modes which have different transmission probabilities from one medium to the other. Total exchanged energy is then obtained by summing the energy of each mode weighed by the mode transmission probability. Maximizing the transfer reduces then to maximizing the transmission probability of the different modes. According to this idea but without performing a detailed optimization study, Ben-Abdallah and Joulain \cite{benabdallah2010} derived with variations calculus a simple analytical condition on Fresnel reflexion coefficients which allows, knowing medium $1$, to determine Fresnel coefficients of the second medium maximizing the transfer. It is then possible to determine the optical properties of both media. It would be interesting to implement this method and compare its results and performances to previously presented methods.
%%***********************************************************************************
%%*********************************Appendices****************************************
%%***********************************************************************************
\section{Appendix : on the calculus of the cutoff wave-vector}
The cutoff wave vector $q_c$ is the upper bound of $q$ that would allow an accurate evaluation of the sum :
\begin{small}
\begin{eqnarray}
\int_{\frac{\omega}{c}}^{\infty} \frac{q}{q_0^2} 
\overbrace{\left(4 \times e^{2  \imath \gamma_3 \delta} 
\frac{{Im}(r_{31}^p) {Im}(r_{32}^p)}{|1-r_{31}^p r_{32}^p e^{2 \imath \gamma_3 \delta} |^2} \right)}^{\tau_{evan}^p(\omega,q)}dq 
& \simeq & 
\int_{\frac{\omega}{c}}^{q_c} \frac{q}{q_0^2} 
\tau_{evan}^p(\omega,q)dq
\label{eq:Qsum}
\end{eqnarray}
\end{small}
According to $\tau_{evan}^p (\omega,q)$ and $(q / q_0) \times \tau_{evan}^p (\omega,q)$ plots (Figures \ref{fig:Chap5-DrudeDispersionTau} and \ref{fig:Chap5-DrudeDispersionTauKappa} respectively. Drude model examples are considered to illustrate the method.), the cutoff wave vector $q_c$ depends on the circular frequency $\omega$. At the present stage and for simplicity sake, we consider a constant cutoff wave vector $q_c$. According to the same figures, the largest wave vectors participating to the transfer are observed for $\omega = \omega_p$. Thus, the constant cutoff wave vector is to be determined at this frequency. Two families of criteria can be considered for $q_c$ definition, whether this latter is based on the transmission coefficient $\tau_{evan}^p (\omega,q)$ or on the transmission coefficient weighted by the normalized wave vector, $(q / q_0) \times \tau_{evan}^p (\omega,q)$.
\subsection{Transmission coefficient criterion}
\begin{figure}[ht]
\begin{center}
\includegraphics[width=0.6\textwidth]{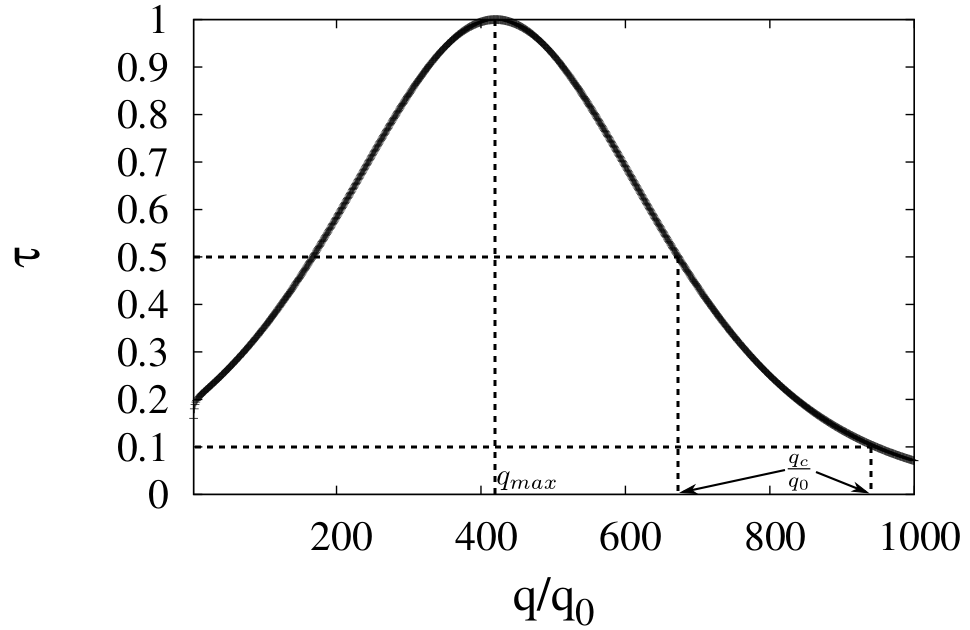}   
\caption{Cutoff wave vector definition based on modes transmission probability $\tau(\omega_p,q)$.}
\label{fig:CutoffTau01}
\end{center}
\end{figure}
A first criterion based on $\tau_{evan}^p (\omega,q)$ can be considered. According to figure \ref{fig:CutoffTau01}, $\tau_{evan}^p (\omega_p,q)$ increases with increasing $\frac{q}{q_0}$ to reach its maximal value $\simeq 1$ at a certain wave vector $q_{max}$ and slowly goes to zero after that. $q_c$ can be defined as the smallest wave vector larger than $q_{max}$ which separates transmitted modes from those with sufficiently small transmission probability, defined by an arbitrary threshold $\tau_{min}$. Then $q_c$ is defined by :
\begin{eqnarray}
\left\{
    \begin{array}{ll}
        q_c \geq q_{max}\\
        \tau_{evan}^p (\omega_p,q_c) = \tau_{min}
    \end{array}
\right.
\label{eq:QcTauDef}
\end{eqnarray}
If we consider a threshold transmission probability $\tau_{min} = 0.5$ for example (this threshold value separates modes that are more likely to be transmitted from those who are not), this leads to $q_c = 2.73 / \delta$ and $q_c = 2.37 / \delta$ for the exact and the approximate optima respectively. A lower threshold of $\tau_{min} = 0.1$ leads to $q_c = 3.55 / \delta$ and $q_c = 3.22 / \delta$ respectively. In all cases, $q_c$ is of the order of $1 / \delta$.
\subsection{Weighted transmission coefficient criteria}
Two criteria have been considered : a direct one that can be directly verified on $(q / q_0) \times \tau_{evan}^p (\omega,q)$ color maps and an indirect one that needs further calculations.
\subsubsection{Full width at half maximum (FWHM)}
\begin{figure}[ht]
\begin{center}
\includegraphics[width=0.6\textwidth]{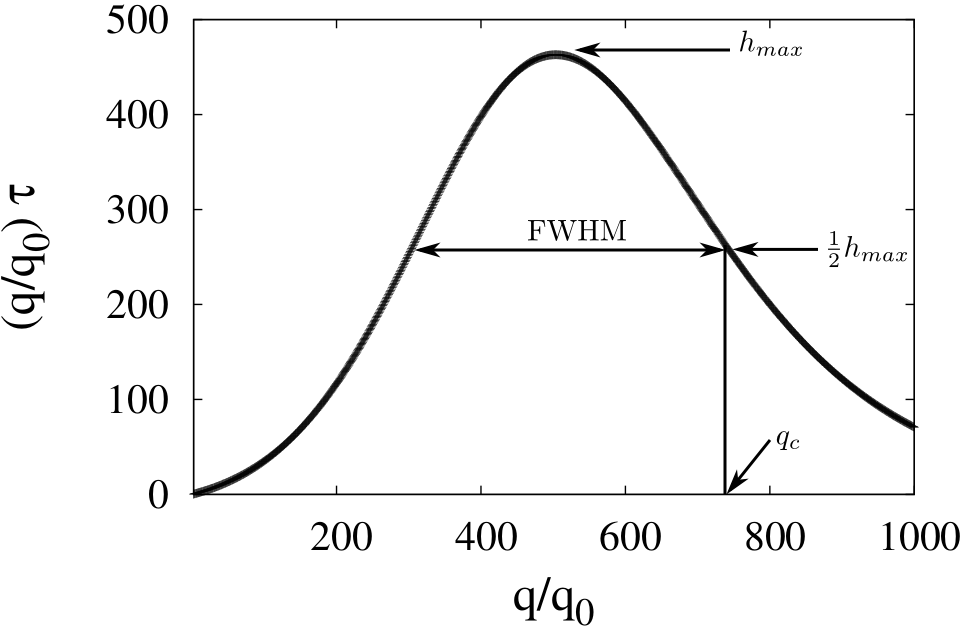}   
\caption{Cutoff wave vector definition based on the monochromatic NF RHF density peak full width at half maximum at $\omega = \omega_p$.}
\label{fig:CutoffFWHM}
\end{center}
\end{figure}

~~\\
Figure \ref{fig:CutoffFWHM} presents the integrand $(q / q_0) \times \tau_{evan}^p (\omega_p,q)$ for $\omega = \omega_p$. A peak is observed. If we consider that the monochromatic flux density at $\omega = \omega_p$ is transmitted by modes with $q$ in a wave vector range equal to the FWHM around the peak, then $q_c$ is the upper bound of the FWHM, i.e. the largest wave vector verifying :
\begin{eqnarray}
\frac{q_c}{q_0} \tau(\omega_p,q_c) = \frac{1}{2} \left[ \frac{q}{q_0} \tau(\omega_p,q) \right]_{max}
\label{eq:QcFWHM}
\end{eqnarray}
For Drude model for instance, this definition led to values of $q_c = 3 / \delta$ and $q_c = 2.7 / \delta$ for the exact and approximate optima respectively.
This criterion is interesting since it can be directly verified on $\frac{q}{q_0}\tau(\omega,q)$ color maps and was used to obtain results reported in the present paper.
\subsubsection{Fractional monochromatic flux}
\begin{figure}[ht]
\begin{center}
\includegraphics[width=0.6\textwidth]{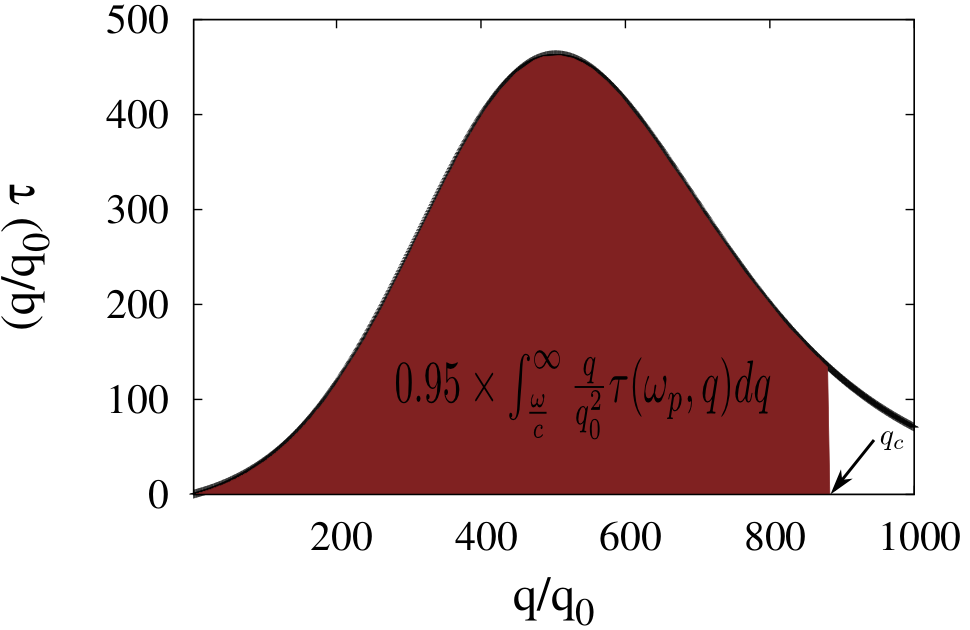}   
\caption{Cutoff wave vector definition according to the fractional monochromatic NF RHF density carried by modes in the range $q \in [q_0,q_c]$.}
\label{fig:Cutoff095}
\end{center}
\end{figure}

~~\\
A second criterion, more complicated to implement though since it can not be read on $\frac{q}{q_0}\tau(\omega,q)$ maps and needs an integral calculation, considers the monochromatic radiative heat flux fraction transmitted in a certain wave vector range.\\
We can then define $q_c$ as the wave vector that verifies (see Figure \ref{fig:Cutoff095}) :
\begin{eqnarray}
\int_{\frac{\omega}{c}}^{q_c} \frac{q}{q_0^2} \tau(\omega_p,q) dq = x_F \times  \int_{\frac{\omega}{c}}^{\infty} \frac{q}{q_0^2} \tau(\omega_p,q) dq 
\label{eq:Qc095}
\end{eqnarray}
where $x_F$ is the monochromatic flux density transmitted fraction.
This second criterion is expected to be more accurate than previously presented ones since it provides a rigorous quantitative information, $x_F$ in this case. 
For $x_F = 0.95$ for example, it leads to $q_c = 3.6 / \delta$ and $q_c \simeq 3 / \delta$ for the exact and the approximate optima respectively. The correction compared to the previous criterion results varies from $20 \%$ to $10 \%$ respectively. However, $q_c$ is still of the order of $1 / \delta$.
%%***********************************************************************************
%%*********************************Acknowledgments***********************************
%%***********************************************************************************
\section*{Acknowledgments}
Authors would like to thank Philippe Ben-Abdallah and Carsten Henkel for fruitful discussions and gratefully acknowledge the support of the Agence Nationale de la Recherche through the Source-TPV Project No. ANR 2010 BLAN 0928 01. 
%%***********************************************************************************
%%*********************************References****************************************
%%***********************************************************************************
%bibliography
%\vspace{-40pt} % this line is just to remove unnecessary white space. I f you do not have any white space in your manuscript, you can remove it.
\section*{References}
% use ieee style
\bibliographystyle{unsrt}
\bibliography{OnMaximalNF-RHT}
%\end{multicols}
\end{document}